\documentclass[prd,twocolumn,showpacs,preprintnumbers,amsmath,amssymb]{revtex4}

\usepackage{graphicx}
\usepackage{epsfig}
\usepackage{dcolumn} 
\usepackage{bm}      

 \newcommand{\zr}[1]{\mbox{\hspace*{#1em}}}

 \newcommand{\RR}{\mbox{\zr{0.1}\rule{0.04em}{1.6ex}\zr{-0.05}{\sf R}}}

\newcommand{\smallfrac}[2]{\mbox{\small ${\displaystyle \frac{#1}{#2}}$}}

\begin{document}
       
\preprint{\sf ADP-07-15/T655}                               
   
\title{Decontracted double BRST on the lattice}

 \author{L.~von~Smekal}
 \author{M.~Ghiotti}\thanks{Address after Sept.~2008: 
 Centre for Computational Finance and\\
  \hspace*{1pt} Economic Agents, 
  University of Essex, Wivenhoe Park, Colchester\\
    \hspace*{1pt}  CO4 3SQ, UK.}
 \author{A.~G.~Williams}

 \affiliation{Centre for the Subatomic Structure of Matter,
   The University of Adelaide,\\
       South Australia 5005, Australia}
                                                                                
\date{July 3, 2008}

\begin{abstract}
We present the Curci-Ferrari model on the lattice. In the massless case 
the topological interpretation of this model with its double BRST symmetry 
relates to the Neuberger 0/0 problem which we extend to include the
ghost/anti-ghost symmetric formulation of the non-linear covariant
Curci-Ferrari gauges on the lattice. The introduction of a
Curci-Ferrari mass term, however, serves to regulate the 0/0
indeterminate form of physical observables  observed by
Neuberger. While such a mass $m$ decontracts the double 
BRST/anti-BRST algebra, which is well-known to result in a loss
of unitarity, observables can be meaningfully defined in the limit 
$m \to 0$ via l'Hospital's rule. At finite $m$ the topological nature
of the partition function used as the gauge fixing device seems
lost. We discuss the gauge parameter $\xi$ and mass $m$ dependence of
the model and show how both cancel when $m\equiv m(\xi)$ is
appropriately adjusted with $\xi $.  
\end{abstract}


                                                                               
\maketitle

\section{Introduction}

In the covariant continuum formulation of gauge theories, in terms of
local field systems, one has to deal with the redundant degrees of
freedom due to gauge invariance. Within the language of local quantum
field theory, the machinery for that is based on the so-called
Becchi-Rouet-Stora-Tyutin (BRST) symmetry which is a global symmetry
and can be considered the quantum version of local gauge
invariance \cite{Nakanishi1990,Alkofer2001}. In short, one starts out
from the representations of a BRST algebra on indefinite metric spaces
with assuming the existence (and completeness) of a nilpotent BRST
charge $Q_B$. The physical Hilbert space can then be defined as the
equivalence classes of BRST closed (which are annihilated by $Q_B$)
modulo exact states (which are BRST variations of others). In QED this
machinery reduces to the usual Gupta-Bleuler construction. For the
generalization thereof, in non-Abelian gauge theories, all is well in
perturbation theory also. Beyond perturbation theory, however, there
is a problem with such a construction that has not been fully and
comprehensively addressed as yet. It relates to the famous Gribov
ambiguity \cite{Gribov1978}, the existence of so-called Gribov copies
that satisfy the Lorenz condition \cite{Jackson2001} 
(or any other local gauge fixing condition) but are related by gauge
transformations, and are thus physically equivalent. As a result of
this ambiguity, the usual definitions of a BRST charge fail to be
globally valid.   

A rigorous non-perturbative framework is provided by lattice gauge
theory. Its strength and beauty derives from the fact that
gauge-fixing is not required. However, in order to arrive at a
non-perturbative definition of non-Abelian gauge theories in the
continuum, from a lattice formulation, we need to be able to perform
the continuum limit in a formally watertight way. And there is the gap
in our present understanding. The same problem as described above
comes back to haunt us in another dress when attempting to fix a gauge
via BRST formulations on the lattice. There it is known as the
Neuberger problem which asserts that the expectation value of any
gauge invariant (and thus physical) observable in a lattice BRST
formulation will always be of the indefinite form 0/0
\cite{Neuberger1987}. 

The BRST algebra requires the introduction of further unphysical
degrees of freedom. These are the Faddeev-Popov ghosts and anti-ghosts
which violate the Spin-Statistics Theorem of local quantum field
theory on positive definite metric (Hilbert) spaces. Contrary to what
the name anti-ghost might suggest, however, in the usual linear
covariant gauges the treatment of ghosts and anti-ghosts is completely
asymmetric. On the other hand, it is also known for many years that it
is possible to extend the BRST algebra to be entirely symmetric
w.r.t. ghosts and anti-ghosts. This additional symmetry arises
naturally in the Landau gauge but can also be extended to more general
gauges, the so-called Curci-Ferrari gauges, at the expense of quartic
ghost self-interactions. The most interesting feature of these gauges
four our purpose, however, is that they allow the introduction of a
mass term for gluons and ghosts \cite{Curci1976}. While such a
Curci-Ferrari mass $m$ breaks the nilpotency of the BRST and anti-BRST
charges, which is known to result in a loss of unitarity
\cite{Ojima1982,deBoer1996} and which therefore meant that
this relatively old model received little attention for many years, it
also serves to regulate the Neuberger zeroes in a lattice formulation. 
In \cite{Kalloniatis2005} this was exemplified in a simple Abelian toy-model
where the zeroes in the numerator and denominator of expectation
values become proportional to $m^2$ and allow to compute a 
finite value for $m^2\to  0 $ via l'Hospital's rule. 

For the $SU(N)$ gauge theory on a finite four-dimensional lattice things
are naturally much more complicated than in the toy model. In this
paper we develop a full lattice formulation of the time-honored
model by Curci and Ferrari with its decontracted double BRST/anti-BRST and
ghost-mass term, as announced in \cite{Ghiotti2006}. After introducing
the general setup for double BRST on the lattice in Sec.~II, we next review
Neuberger's no-go-theorem in a generalized version to include the
ghost/anti-ghost symmetric case of the non-linear covariant
Curci-Ferrari gauges for $m^2=0$ in Sec.~III, a case originally
excluded by Neuberger. At non-vanishing 
Curci-Ferrari mass the partition function of the model used as the
gauge-fixing device is shown to be polynomial in $m^2$
and to be thus non-vanishing, in a special gauge-parameter limit in
Sec.~IV. In this way regularizing the Neuberger zeroes, the leading
power of that polynomial can be extracted from a suitable number of
derivatives (w.r.t.~$m^2$) before the limit $m^2\to 0$ is taken, in
the spirit of l'Hospital's rule. This could provide a lattice BRST
model without Neuberger problem. The massive Curci-Ferrari model is
no-longer purely topological in nature, however, and as a result, its
gauge-parameter $\xi $ independence requires tuning of
the Curci-Ferrari mass with $\xi$  as explained in Sec.~V. The
gauge-orbit independence of this procedure is discussed in Sec.~VI.
A short summary is given in Sec.~VII, and our conclusions and outlook 
are provided in Sec.~VIII.


\section{Double BRST on the lattice}
\label{DBL}

For the topological lattice formulation of the double BRST symmetry of
the ghost/anti-ghost symmetric covariant gauges we start out from the
standard gauge-fixing functional $V_U[g]$ of covariant gauges which
here assumes the role of a Morse potential on a gauge orbit,
\begin{equation}
  V_U[g] \, = \, - \smallfrac{1}{2\rho} \, \sum_i \sum_{j\sim i} \mbox{tr}\,
  U^g_{ij} \, = \, - \smallfrac{1}{\rho} \sum_{x,\, \mu} \mbox{Re tr}\,
  U_{x,\mu}^g \; . 
\end{equation}
Here, in the first form, $U_{ij} \in SU(N)$ is the directed link variable 
connecting nearest neighbor sites $i$ and $j$.  The sum $j\sim i$
denotes summation over all nearest neighbors $j$ of site $i$. We assume
periodic boundary conditions.  
The double sum thus runs twice over all links $\langle ij\rangle$, and
with $U_{ij}^\dagger = U_{ji}$ it is therefore equivalent to the
simple sum over links in the second form, where $U_{x,\mu} $ stands
for the same link field $U$ at position $x$ in direction $\mu$. The
constant $\rho$ is the normalization of the $SU(N)$ generators $X$. We use
anti-Hermitian   $ [ X^a,X^b] = f^{abc} X^c$ with tr$X^a X^b= -\rho \,
\delta^{ab}$. We explicitly only need the fundamental representation,
where $\rho=\rho_{\mbox{\tiny fund}} = 1/2$.

As usual, under gauge transformations the link variables $U$ transform
\begin{equation} 
   U_{ij} \, \to \,  U_{ij}^g \, = \, g_i^\dagger U_{ij} g_j \; .
\end{equation}
BRST transformations $s$ and anti-BRST transformations $\bar s$ in the
topological setting do not act on the link variables $U$ directly,
but on the gauge transformations $g_i$ like infinitesimal 
right translations in the gauge group 
with real ghost and anti-ghost Grassmann fields
$c_i^a$, $\bar c_i^a$ as parameters, respectively,
\begin{equation}
   s g \, = \, g \, X^a c^a \, = \, g c  \; , \;\;    
   \bar s g  \, = \, g \, X^a \bar c^a \, = \, g \bar c \; , 
\end{equation}
where we introduced Lie-algebra valued, anti-Hermitian ghost
fields $c_i \equiv X^a c_i^a$ with $c_i^\dagger = - c_i$,
and analogous anti-ghost fields $\bar c_i\equiv X^a \bar c_i^a$.
For consistency, we furthermore require
\begin{equation}
   s g^\dagger \, = \, (s g)^\dagger \, = \, - c g^\dagger  \; , \;\;    
   \bar s g^\dagger  \, = \, (\bar s g)^\dagger \, = \, - \bar c
   g^\dagger 
\; . 
\end{equation}
For the gauge-transformed link variables this then implies
\begin{equation} 
s U^g_{ij} \, = \, - c_i U_{ij}^g + U_{ij}^g c_j \; , \;\; 
\bar s U^g_{ij} \, = \, - \bar c_i U_{ij}^g + U_{ij}^g \bar c_j \; .
\label{BRSUg}
\end{equation}

\vspace*{-1pt}
\noindent
The BRST transformations for (anti)ghosts and Naka\-nishi\--Lautrup fields $b$
are straightforward lattice analogues (per site) of their continuum
counterparts, see, {\it e.g.}, Refs.~\cite{Baulieu1982,Thierry-Mieg1985},
\begin{eqnarray}
  s c^a &=& - \smallfrac{1}{2} (c \times c)^a  \; , \label{BRSc} \\
  s \bar c^a &=& b^a - \smallfrac{1}{2} (\bar c \times c)^a \; ,\label{BRScb}\\
  s b^a &=&  - \smallfrac{1}{2} (c \times b)^a  
 - \smallfrac{1}{8} \big((c \times c) \times \bar c  \big)^a \;
  . \label{BRSb}  
\end{eqnarray}
The relatively obvious notation of using the ``cross-product'' 
herein refers to the structure constants for $SU(N)$, for example,
$(\bar c \times c)^a \equiv f^{abc} \bar c^b c^c$.

In the ghost/anti-ghost symmetric gauges as considered here, the
anti-BRST variations are obtained by  substituting $c \to \bar c$ and
$\bar c \to - c$ according to Faddeev-Popov conjugation. Thus,
\begin{eqnarray}
  \bar s c^a &=& - b^a - \smallfrac{1}{2} (\bar c \times c)^a \; ,
    \label{aBRSc} \\
  \bar s \bar c^a &=&  - \smallfrac{1}{2} (\bar c \times \bar c)^a  \;
  ,  \label{aBRScb} \\
  \bar s b^a &=&  - \smallfrac{1}{2} (\bar c \times b)^a  
 + \smallfrac{1}{8} \big((\bar c \times \bar c) \times c  \big)^a \; .
  \label{aBRSb} 
\end{eqnarray}
The action of the topological lattice model for gauge fixing a la
Faddeev-Popov with double BRST invariance can then be written in
compact form as
\begin{equation} 
S_{\mbox{\tiny GF}} \, = \, i \, s \bar s \, \Big( V_U[g] + i
\smallfrac{\xi}{2\rho}  
\sum_i \mbox{tr}\, \bar c_i c_i \Big) \; . \label{S_GF}
\end{equation}
This is the lattice counterpart of the continuum gauge-fixing 
Lagrangian
\begin{equation}
 \mathcal{L}_{\mbox{\tiny GF}} \, = \, \smallfrac{i}{2} \, s \bar s
 \big( A_\mu^a A_\mu^a  - i \xi \bar c^a c^a \big)\;\; \mbox{with} \;\; 
S_{\mbox{\tiny GF}} \, =  \int d^Dx \,  \mathcal{L}_{\mbox{\tiny GF}}
 \label{L_GFcont}
\end{equation}
in $D$ Euclidean dimensions.

For the purpose of a self-contained presentation we work out the
double (anti-)BRST variation on the right of (\ref{S_GF}) explicitly in
Appendix~\ref{expl_ders}. This leads to 
\begin{eqnarray}
S_{\mbox{\tiny GF}} &=& \sum_i \,  \Big\{ - i b^a_i  F_i^a(U^g)
\, - {i}\,\bar c^a_i  {M_{\mbox{\tiny FP}}}_i^a[c]  \label{SGFexpl}\\
&& \hskip 2cm +\, 
\smallfrac{\xi}{2} b_i^a b_i^a \, + \, \smallfrac{\xi}{8}\, 
(\bar c_i \times c_i)^2 \, \Big\} \; , \nonumber
\end{eqnarray}
where 
\begin{equation} 
 F^a_i(U^g) \, = \,  -\smallfrac{1}{2\rho} \,  \sum_{j\sim i}
 \mbox{tr}\,\big( X^a(  U^g_{ij} -   U^g_{ji}) \big) 
\label{std-gfc}
\end{equation}
defines, of course, the standard gauge-fixing form of covariant gauges
with the continuum limit,
\begin{equation} 
 F^a_i(U^g) \, \stackrel{a\to 0}{\longrightarrow} 
 \,  a^2\, \partial_\mu {A_\mu^g}^a  \, +\, \mathcal{O}(a^4) \; .
\end{equation}
The Faddeev-Popov operator ${M_{\mbox{\tiny FP}}}^{ab}_{ij}$ is
obtained from the short-hand notation in (\ref{SGFexpl}),
\begin{equation}
 \sum_{i} \bar c_i^a \,
     {M_{\mbox{\tiny FP}}}^{a}_{i}[c]  \, = \, 
 \sum_{i,\,j} \bar c_i^a \,
     {M_{\mbox{\tiny FP}}}^{ab}_{ij} \,  c_j^b \, ,
\end{equation}
and given explicitly for later reference 
in alternative forms in Eqs.~(\ref{edBRSFP}) or (\ref{sBRSTFP}). 
It is symmetric w.r.t.~simlutaneous interchanges of color and site
indices, and identical to the one obtained in \cite{Zwanziger1994} as
the Hessian of $V_U[g]$ from variations along one-parameter subgroups
of the $SU(N)$ gauge group. In the continuum limit it reduces to 
the symmetrized and thus Hermitian
\begin{equation} 
 {M_{\mbox{\tiny FP}}}^{ab}_{ij} \, \stackrel{a\to 0}{\longrightarrow} 
 \,  - a^2\smallfrac{1}{2} \,\big(\partial  D^{ab} + D^{ab}
 \partial \big) \, \delta(x-y)  \, +\, \mathcal{O}(a^4) \; \nonumber
\end{equation}   
of the ghost/anti-ghost symmetric Curci-Ferrari gauges. In
contrast, the Faddeev-Popov operator of the linear covariant gauges
for $\xi\not=0$ is not a Hessian because it is not symmetric. It can
be read off as a byproduct of our BRST derivation from Eq.~(\ref{stdFP_2}).
In particular, this non-symmetric Faddeev-Popov operator needs to be
used when implementing other linear-covariant gauges such as the
Feynman gauge with $\xi = 1$ on the lattice as discussed in
\cite{Giusti1996,Giusti1999}.  
In Landau gauge $\xi=0$ the distinction is an illusion. To keep the symmetric
Hessian for $\xi\not=0$, however, is only possible within the
ghost/anti-ghost symmetric framework where it necessarily comes along with the
quartic ghost self-interactions in (\ref{SGFexpl}).
 
The full symmetry of the ghost/anti-ghost symmetric Curci-Ferrari
gauges \cite{Curci1976,Thierry-Mieg1985} 
is given by a semidirect product of a global $SL(2,\RR)$, which
includes ghost number and Faddeev-Popov conjugation, with
the BRST/anti-BRST symmetries as used above \footnote{Also see Appendix A of
  Ref.~\cite{Alkofer2001}.}.  
This is the global symmetry of the Landau gauge, and it is sometimes 
referred to as extended BRST symmetry, see~\cite{Nakanishi1990}. 

Among the general class of all
covariant gauges \cite{Baulieu1982}, with a Lagrangian which is 
polynomial in the fields, Lorentz, globally gauge and BRST invariant,
and renormalizable in $D=4$, the ghost/anti-ghost symmetric case is
special and interesting in that it allows to smoothly connect to the
Landau gauge for $\xi \to 0$, without changing the global symmetry
properties.    

In particular, introducing with \cite{Baulieu1982} a second gauge
parameter $\beta \in [0,\, 1] $, to interpolate between the various
generalized covariant gauges,  the linear covariant gauges of standard 
Faddeev-Popov theory correspond to the line $\beta = 0$ in the two
gauge-parameter plane $(\xi,\,\beta)$. Along this line, the global
symmetry changes abruptly when reaching the Landau gauge limit; and for
$\beta = 1$, one obtains a mirror image of standard Faddeev-Popov 
theory with the roles of ghosts and anti-ghosts interchanged. The
ghost/anti-ghost symmetric gauges discussed here 
then correspond to the line $\beta = 1/2$. The $\xi = 0$ gauge is
$\beta$-independent. The whole interval for $\beta \in [0,1]$ 
at $\xi =0$ is equivalent and corresponds to the Landau gauge. The important
difference is, however, that the $SL(2,\RR) $ symmetric line at
$\beta = 1/2$ provides a unique class of covariant gauges which share the
full extended BRST symmetry of the Landau gauge for any value of
$\xi$. The limit $\xi \to 0$ is thus a smooth one, as far as this
symmetry is concerned, only along the line of $\beta=1/2$. The price to pay are
the quartic ghost self-interactions in (\ref{SGFexpl}) which again
vanish only in the Landau gauge limit. 

For a further discussion of the general ghost creating gauges, and their
geometrical interpretation, see \cite{Thierry-Mieg1985}. The one-loop
renormalization was first discussed in \cite{Baulieu1982}, for
explicit calculations of renormalization constants and anomalous
dimensions of the ghost/anti-ghost symmetric case up to including
the three-loop level, see \cite{deBoer1996,Gracey2002}.
The Dyson-Schwinger equations of these gauges were studied in
\cite{Alkofer2003}. A non-renormalization theorem relating to the
Curci-Ferrari mass was recently reported in \cite{Wschebor2007}.

\section{The Neuberger problem}

Following Neuberger, we introduce an auxiliary parameter $t$ in the 
Euclidean partition function to be used as the gauge-fixing device via the
Faddeev-Popov procedure of inserting unity into the unfixed partition
function of $SU(N)$ lattice gauge theory. The gauge-fixing action of 
the double BRST invariant model given by (\ref{S_GF}) consists of two
terms both of which are separately BRST (and anti-BRST)
exact. Multiplying the $1^\mathrm{st}$ term in (\ref{S_GF}) by the
real parameter $t$ amounts to a mere redefinition of the Morse
potential which should have no further effect. We can therefore write the
gauge-fixing partition function with double BRST,
\begin{eqnarray}
Z_{\mbox{\tiny GF}}(t) &=& \int d[g, b,\bar c, c]  \, \times
\label{Z_GF}\\
&& \hskip .5cm  \exp\Big\{-  i s \bar s
\Big( t \, V_U[g] \, + \,  i\smallfrac{\xi}{2\rho} 
 \sum_i \mbox{tr}\, \bar c_i c_i \Big)  
\Big\} \; , \nonumber
\end{eqnarray}
which is independent of the set of link variables $\{U\}$ and the
gauge parameter $\xi$ because of its topological nature. Moreover, the $t$
independence is really no different from the $\xi$ independence here,
and it is thus rather obvious. Explicitly, the derivative with respect
to $t$ (or $\xi$) produces the expectation value of a BRST exact
operator which vanishes, {\it i.e.}, 
\begin{equation}
   Z'_{\mbox{\tiny GF}}(t) \, = \, 0\;.  \label{t-indep}
\end{equation}
At $t\!=\!0$ on the other hand, we obtain with the BRST variations
given in (\ref{sbcbc}) and (\ref{scbb}) of Appendix~\ref{expl_ders},   
\begin{eqnarray} 
\label{GaussBonnet}
Z_{\mbox{\tiny GF}}(0) &=& \mathcal{N} \int d[b,\bar c, c]  \, \times \\
&& \hskip .5cm   \exp\Big\{-  \sum_i \,  \left( 
\smallfrac{\xi}{2} b_i^a b_i^a \, + \, \smallfrac{\xi}{8}\, 
(\bar c_i \times c_i)^2 \right) \,
\Big\} \; , \nonumber
\end{eqnarray}
where the volume of the gauge group on the lattice, from
the invariant integrations $\prod_i dg_i$ via the Haar measure over
$g_i\in SU(N)$ per site $i$, is absorbed in the constant $\mathcal{N}$.
The Gaussian integrations over the Nakanishi-Lautrup fields $b$ are also
well-defined and produce a factor $(2\pi/\xi)^{(N^2\!-\!1)/2}$ per site.

One might be tempted to conclude at this point that the quartic ghost
self-interactions in (\ref{GaussBonnet}) might remove the
uncompensated Grassmann integrations of the linear covariant gauges
where no such self-interactions occur. The ghost/anti-ghost 
integrations at $t=0$ also factorize into independent 
integrations $d\bar c^a_i d c^a_i$  over $2(N^2-1)$ Grassmann
variables per site. For $N=3$, for example, the $4^\mathrm{th}$ order
term of the exponential in (\ref{GaussBonnet}) produces a monomial in
$\bar c^a_i$, $c^a_i$ which contains each of these $16$ Grassmann
variables exactly once, so that their integration might produce 
a non-vanishing result. This is not the case, however. Working out the 
prefactor of this monomial, as we will do explicitly in the more
general case with including a non-vanishing Curci-Ferrari mass $m$
below, one finds that the prefactor of this term in
(\ref{GaussBonnet}) vanishes in the massless case
and thus,
\begin{equation}
  Z_{\mbox{\tiny GF}}(0) \, = \, 0\;. \label{Z-zero}
\end{equation}
Because of the $t$-independence (\ref{t-indep}), this implies the
vanishing of the gauge-fixing partition function (\ref{Z_GF}) of the
ghost/anti-ghost or $SL(2,\RR)$ symmetric formulation with double BRST
invariance in the same way as that of standard Faddeev-Popov theory 
observed in \cite{Neuberger1987}. As for the latter, the sign-weighted
sum over all Gribov copies, as originally proposed to generalize the
Faddeev-Popov procedure in presence of Gribov copies
\cite{Hirschfeld1979,Fujikawa1979}, vanishes. 
 
This cancellation of Gribov copies is well-known \cite{Sharpe1984}. 
The fact that it also arises here, in the ghost/anti-ghost symmetric
formulation with its quartic self-interactions, directly relates to the
topological interpretation \cite{Baulieu1998,Schaden1998} 
of the Neuberger zero: $Z_{\mbox{\tiny GF}}$ can be viewed as the
partition function of a Witten-type topological model 
to compute the  Euler characteristic $\chi$  of the gauge
group. On the lattice the gauge group is a direct product of $SU(N)$'s
per site, and because the Euler characteristic factorises,
\[
Z_{\mbox{\tiny GF}} = \chi(SU(N)^{\#{\rm sites}}) =
\chi(SU(N))^{\#{\rm sites}} = 0^{\# {\rm sites}}\; . 
\]
For $t=0$ the action in (\ref{Z_GF}) decouples from the link-field
configuration and $Z_{\mbox{\tiny GF}}(0) $, albeit computing the same
topological invariant, has of course no effect in terms of fixing a
gauge. In the present formulation, with $Z_{\mbox{\tiny GF}}(0) $ in
(\ref{GaussBonnet}), the independent Grassmann integrations per site
of the quartic-ghost term which contains the curvature of $SU(N)$ each 
compute its Euler characteristic via the Gauss-Bonnet theorem
\cite{Birmingham1991}. This explicitly produces one factor of zero per
site on the lattice. And it provides the topological explanation for the 
vanishing of the prefactor of the corresponding monomial of degree
$2(N^2-1)$ in the Grassmann variables  $\bar c$, $c$, which could
otherwise exist in the expansion of the exponential in
(\ref{GaussBonnet}) for all odd $N$. For $N=3$, for example, the  
zero in this prefactor arises, upon normalordering, from a cancellation of     
368 non-vanishing individual terms when expanding the square of the
square of the quartic ghost self-interaction. This cancellation
would be rather unnatural to arise accidentally, without such explanation.

The vanishing of the gauge-fixing partition function
at $t=0$ part in Neuberger's argument, in the ghost/anti-ghost
symmetric gauges with their $SL(2,\RR) \rtimes $ double BRST symmetry,
therefore most directly reflects the topological origin of the
Neuberger zero. Eq.~(\ref{GaussBonnet}) precisely represents a product
of one Gauss-Bonnet integral expression for $\chi(SU(N))$ per
site of the lattice. 

Note that the gauge parameter $\xi$  can be removed completely from 
the expression for $Z_{\mbox{\tiny GF}}(0)$ in
Eq.~(\ref{GaussBonnet}) by a rescaling $\sqrt{\xi}\,  b \to b$ and
$\sqrt[4]{\xi}\, \bar c \to \bar c$, $\sqrt[4]{\xi}\, c \to c$, which
leaves the integration measure unchanged. The same rescaling for the
full gauge-fixing partition function $Z_{\mbox{\tiny GF}}(t)$ in
(\ref{Z_GF}), which amounts to replacing the action in $S_{\mbox{\tiny
    GF}}$ in (\ref{SGFexpl}) by
\begin{eqnarray}
S_{\mbox{\tiny GF}}(t) &=& \sum_i \,  \Big\{ - i t  b^a_i  F_i^a(U^g)
\, - {i} t \,\bar c^a_i  {M_{\mbox{\tiny FP}}}_i^a[c]
 \label{SGFtexpl}\\ 
&& \hskip 2cm +\, 
\smallfrac{\xi}{2} b_i^a b_i^a \, + \, \smallfrac{\xi}{8}\, 
(\bar c_i \times c_i)^2 \, \Big\} \; , \nonumber
\end{eqnarray}
furthermore shows that $t$ and $\xi$ really represent a single
parameter $t/\sqrt{\xi} $. Setting $t=0$ in Neuberger's argument
is therefore the same as the $\xi \to \infty $ limit which is usually
what is considered as the Gauss-Bonnet limit in topological quantum
field theory \cite{Birmingham1991}.  As mentioned above, there is no
gauge-fixing in this limit, but it provides a simple way to compute
the value (zero here) of the partition function which is independent
of $t/\sqrt{\xi}$. 

In the opposite limit, that of the Landau gauge $\xi \to 0$ or
$t/\sqrt{\xi} \to \infty $, of course, $Z_{\mbox{\tiny GF}}(t)$ still
reduces to the sign-weighted sum over all Gribov copies as usual
\cite{Hirschfeld1979,Fujikawa1979},   
\begin{equation} 
  Z_{\mbox{\tiny GF}}(t) \to \sum_{\mbox{\tiny copies $\{g^{(i)}\!\}$}} 
  \mbox{sign} \,\big( \mbox{det} \, M_{\mbox{\tiny FP}}(U^{g^{(i)}})
  \big) \; ,
  \label{P-H}
\end{equation}
which because of the $t$ (and $\xi$) independence (\ref{t-indep})
thus computes the same topological zero
\cite{Sharpe1984,Baulieu1998,Schaden1998}, in this case via the 
Poincar\'e-Hopf theorem \cite{Birmingham1991}. 

\bigskip

\section{The massive Curci-Ferrari model on the lattice}

In the previous section we have seen that the quartic ghost
self-interactions of the  $SL(2,\RR) \rtimes $ double BRST symmetric
Curci-Ferrari gauges have no effect on the disastrous conclusion of the
0/0 problem in lattice BRST. They rather serve to reveal most clearly 
the topological origin of this problem. 

We will demonstrate explicitly below that this zero can be
regularized, however, by introducing a Curci-Ferrari mass $m$, as
proposed in \cite{Kalloniatis2005,Ghiotti2006}. The gauge-fixing 
action $S_{\mbox{\tiny GF}}$ is thereby once more replaced by 
\begin{eqnarray}
\label{S_mGF}
S_{\mbox{\tiny mGF}}(t) \, = \, i \,( s \bar s-im^2) \, \Big( t\, 
 V_U[g] \, + \, i\xi \sum_i \mbox{tr}\, \bar c_i c_i \Big) 
\end{eqnarray}
(where we dropped in the $2^\mathrm{nd}$ term 
the factor $1/(2\rho) = 1$, in the  fundamental representation).   
The BSRT and anti-BRST transformations of $U^g$, $\bar c $ and $c$ in
Eqs.~(\ref{BRSUg}), (\ref{BRSc}), (\ref{BRScb}) and (\ref{aBRSc}),
(\ref{aBRScb}) of Sect.~\ref{DBL} remain unchanged. Those for the
Nakanishi-Lautrup $b$-fields, Eqs.~(\ref{BRSb}) and (\ref{aBRSb}), are
replaced by \cite{Thierry-Mieg1985},
\begin{eqnarray}
  s b^a &=& im^2 \, c^a \, - \smallfrac{1}{2} (c \times b)^a  
 - \smallfrac{1}{8} \big((c \times c) \times \bar c  \big)^a \;
  , \label{mBRSb}  \\
  \bar s b^a &=& im^2 \, \bar c^a \, - \smallfrac{1}{2} (\bar c \times b)^a  
 + \smallfrac{1}{8} \big((\bar c \times \bar c) \times c  \big)^a \; .
  \label{maBRSb} 
\end{eqnarray}
In the derivation of the explicit form for $S_{\mbox{\tiny mGF}}(t)$,
using these modified (anti-)BRST transformations, the only
modification in comparison to Sect.~\ref{DBL} and
App.~\ref{expl_ders}, arises from $s(\bar c^a 
b^a)$ in (\ref{scbb}), which now becomes,
\begin{equation}
  s \big( \bar c^a b^a \big) \, = -i m^2 \, \bar c^a c^a \,
   + \, b^a b^a \, +\, \smallfrac{1}{4} 
\, (\bar c \times c )^2 \; . \label{mscbb}
\end{equation}
The additional first term on the right contributes an additional
term $-i(\xi/2) m^2 \bar c^a_i c^a_i $ to the gauge-fixing Lagrangian,
{\it c.f.}, Eq.~(\ref{S_GF-standard}). Together with the same
contribution from the explicit mass term $-i(\xi/2) m^2 \bar c^a_i
c^a_i $ in (\ref{S_mGF}) we therefore
obtain {\em twice that} as the final ghost mass-term of the massive
Curci-Ferrari model (this subtlety will be worth remembering for later).    
The action of the massive Curci-Ferrari model therefore becomes,
explicitly,
\begin{eqnarray}
S_{\mbox{\tiny mGF}}(t) &=&  m^2 t\, V_U[g] \, + \label{mSGFtexpl}\\[2pt]
&& \hskip -.4cm \sum_i \,  \Big\{ - i t  b^a_i  F_i^a(U^g)
\, - {i} t \,\bar c^a_i  {M_{\mbox{\tiny FP}}}_i^a[c]
 \nonumber \\[-4pt] 
&& \hskip .6cm +\, 
\smallfrac{\xi}{2} b_i^a b_i^a \, - i m^2 \xi\, \bar c_i^a c_i^a  \, +
\, \smallfrac{\xi}{8}\,  (\bar c_i \times c_i)^2 \, \Big\} \; . \nonumber
\end{eqnarray}
BRST and anti-BRST transformations are no-longer nilpotent at finite
$m^2$, but we have  \cite{Nakanishi1990,Curci1976,Thierry-Mieg1985}
\begin{eqnarray} 
    s^2 = i m^2 \sigma^+ \; , && \bar s^2 = -i m^2 \sigma^- \; ,
            \nonumber \\
            s\bar s + \bar s s &=& -i m^2 \sigma^0 \; , \label{superalgebra}
\end{eqnarray}
where $\sigma^\pm $ and $\sigma^0$ generate the global $SL(2,\RR)$
including ghost number and Faddeev-Popov conjugation. The
Curci-Ferrari mass decontracts the $sl(2,\RR) \rtimes$ double BRST
algebra of the massless case to the $osp(1|2)$ superalgebra extension
of the Lie algebra of the 3-dimensional Lorentz group
$SL(2,\RR)$. Conversely, the $m^2\to 0$ limit   
is interpreted as an Inonu-Wigner contraction of the simple
superalgebra  $osp(1|2)$ \cite{Nakanishi1990,Thierry-Mieg1985}.
The BRST and anti-BRST invariance of the massive Curci-Ferrari action
in (\ref{S_mGF}) itself follows readily from this algebra as given in
(\ref{superalgebra}), noting that only $\bar c$ and $c$ transform
non-trivially under the $SL(2,\RR)$. 
 
We emphasize that this algebra decontraction has from the very
beginning been known to lead to a breakdown of unitarity when
attempting a BRST cohomology construction of a physical Hilbert space
in analogy to the massless case \cite{Curci1976}.
In fact, explicit examples exist for states of negative norm surviving
in any such construction \cite{deBoer1996,Ojima1982}. They do not
belong to BRST quartets and can therefore not be removed by the
quartet mechanism \cite{Nakanishi1990}. Only through the algebra
contraction by $m^2\to 0$ do these states reduce to zero norm components
which have no effect on the physical $S$-matrix elements. 

Here we deliberately do not want to interpret the mass parameter by
Curci and Ferrari as a physical mass. It rather serves to meaningfully
define a limit $m^2\to 0$ on the lattice, perhaps in parallel with the
continuum limit, to recover nilpotent (anti-)BRST transformations. 

To study the parameter dependence, we first define the partition
function of the massive Curci-Ferrari model, explicitly listing all
three parameters (even though these again really only represent 2
independent ones as we will show below),  
\begin{equation} 
Z_{\mbox{\tiny mGF}}(t,\xi,m^2) \,=\, \int d[g, b,\bar c, c]  \, 
 \exp\big\{- S_ {\mbox{\tiny mGF}}(t) \big\} \; ,
\label{Z_mGF}
\end{equation}
with $S_ {\mbox{\tiny mGF}}(t) $ from (\ref{S_mGF}) or (\ref{mSGFtexpl}). 
We note in passing that the terms proportional to $m^2$ in
the massive Curci-Ferrari action (\ref{mSGFtexpl}) are given by  
\begin{equation}
 \mathcal{O} (t,\xi) \equiv  t V_U[g] - i \xi \sum_i \bar c_i^a c_i^a
 \; , \label{LatKondOp}
\end{equation}
or, in the continuum,
\begin{equation}
  \mathcal{O} (t,\xi) =  \int d^Dx  \left( \smallfrac{t}{2} A_\mu^a(x)
  A_\mu^a(x) - i \xi   \bar c^a(x)  c^a(x) \right) \, .
\end{equation}
For $t=1$ this coincides with the on-shell BRST invariant (at $m^2=0$) 
operator proposed by Kondo as a possible candidate for a dimension 2
condensate \cite{Kondo2001}.
The doubling of the explicit ghost mass-term in (\ref{S_GF}), by the BRST
variation of $\bar c b$ in (\ref{mscbb}) as mentioned above, is
crucial here. Without this difference in the relative factor of 2 between the
two terms in $\mathcal O(t,\xi)$ and the gauge fixing functional
\begin{equation}
   -i W_{\mbox{\tiny GF}} 
     =  t V_U[g] -i  \smallfrac{\xi}{2}  \sum_i \bar c_i^a c_i^a \; ,
\end{equation}
one could not have both, the on-shell BRST invariance of $\mathcal O$
and the gauge-fixing action in (\ref{S_GF}) from the double BRST variation $
S_{\mbox{\tiny GF}}  = s \bar s \, W_{\mbox{\tiny GF}} $, at the same time.
 
The observation that the mass terms in (\ref{mSGFtexpl}) are given by
$m^2 \mathcal O(t,\xi) $ could in principle be used to obtain the 
expectation value of Kondo's operator from the derivative 
\begin{equation} 
        \langle \mathcal{O}(t,\xi) \rangle = -\frac{\partial}{\partial
        m^2} \, \ln Z_{\mbox{\tiny mGF}}(t,\xi,m^2) \Big|_{m^2=0} \; ,
\label{mderiv}
\end{equation}
upon insertion into the unfixed partition function of lattice gauge
theory, {\it i.e.}, with taking the additional expectation value in
the gauge-field ensemble. 
As any other observable at $m^2 =0$ this expectation value
as it stands, unfortunately, of course also suffers from Neuberger's
$0/0$ problem of lattice BRST.  

In order to demonstrate that the Curci-Ferrari mass
regulates the Neuberger zero, for $t=0$ we will verify by explicit
calculation that  
\begin{equation}
  Z_{\mbox{\tiny mGF}}(0,\xi,m^2) \, \not= \, 0\;. \label{Zm-nzero} 
\end{equation}
In fact, from (\ref{Z_mGF}), (\ref{mSGFtexpl}),
\begin{eqnarray} 
\label{qGaussBonnet}
Z_{\mbox{\tiny mGF}}(0,\xi,m^2) &=& \mathcal{N} \int d[b,\bar c, c]
\, \times \\ && \hskip -2.5cm   \exp\Big\{-  \sum_i \,  \left( 
\smallfrac{\xi}{2} b_i^a b_i^a \, - i m^2 \xi\, \bar c_i^a c_i^a  \,
+ \, \smallfrac{\xi}{8}\,  
(\bar c_i \times c_i)^2 \right) \,
\Big\} \; , \nonumber
\end{eqnarray}
which again factorises into independent Grassmann (and $b$-field)
integrations per site on the lattice. Using the same 
rescaling $\sqrt{\xi}\,  b \to b$ and $\sqrt[4]{\xi}\, \bar c \to \bar
c$, $\sqrt[4]{\xi}\, c \to c$ as mentioned in the last section, we obtain,
\begin{eqnarray}
  Z_{\mbox{\tiny mGF}}(0,\xi,m^2) & =& \label{Zm} \\
&& \hskip -1cm \left( V_{N}\,
  (2\pi)^{(N^2\!-1)/2}\, I_N\big(m^2\sqrt{\xi}\,\big) \right)^{\# {\rm
      sites}} , 
\nonumber  
\end{eqnarray}

\vspace*{.1cm}

\noindent
where $V_{N}$ is the group volume of $SU(N)$, and 
\begin{eqnarray} 
 I_N(\widehat m^2) & = & \int \prod_{a=1}^{N^2\!-1} d(i\bar c^a) d c^a
 \, \label{cbc-siteint}\\
&& \hskip 1.4cm
 \exp\left\{ i \widehat m^2\,  \bar c\! \cdot \! c \, - \,
 \smallfrac{1}{8}\,   (\bar c \times c)^2 \right\} \; , \nonumber
\end{eqnarray}
where we used the rather obvious abbreviations $\bar c \!\cdot\!c = \bar c^a
c^a $, $(\bar c \times c)^a = f^{abc} \bar c^b c^c$, and $\widehat m^2
= m^2 \sqrt{\xi}$. Note that we define the Grassmann integration
measure to include the imaginary unit $i$ with the real anti-ghosts $\bar c$ so
as to reproduce the result of integrating over complex conjugate 
Grassmann variables $c^a \pm i \bar c^a$. Expanding the exponential
and collecting the relevant powers in the ghost/anti-ghost variables,
for $SU(2)$ we straight-forwardly obtain,
\begin{equation} 
I_2(\widehat m^2) \, = \smallfrac{3}{4} \widehat m^2 \left(
1+\smallfrac{4}{3} \widehat m^4\right) \; . \label{I_2}
\end{equation}
For $SU(3)$ the computation is a bit more tedious, the result is
\begin{equation} 
I_3(\widehat m^2) \, = \smallfrac{45}{64} \widehat m^4 \left(
1+ 4 \widehat m^4 + \smallfrac{64}{15} \widehat m^8 +
\smallfrac{64}{45} \widehat m^{12} \right) \, . \label{I_3}
\end{equation}
In both cases we factorised the leading power for $\widehat m^2\to 0$.
$I_N(\widehat m^2)$ is polynomial in $\widehat m^2 = m^2\sqrt{\xi} $
of degree $N^2\!-1$, for all $N$. The successively lower 
powers of $\widehat m^2$ decrease by 2 in each step 
in this polynomial, reflecting an increasing power of the quartic
ghost self-interactions contributing to each term. Therefore, the
polynomials $I_N(\widehat m^2)$ are odd/even in $\widehat m^2$  for $N$
even/odd.  

Because the polynomial is odd for all even $N$ there can thus not be
an order-zero term in the first place. The powers of the quartic
interactions alone never match the number of independent Grassmann
variables, and the Neuberger zero at $\widehat m^2=0$ arises rather 
trivially  for even $N$, for the same reason that the Euler characteristic
of an odd-dimensional manifold, here of dimension $N^2\!-\!1$, 
necessarily vanishes.

For $N$ odd, $I_N(\widehat m^2) $ is an even polynomial which could in
principle have an order zero, constant term. The fact that this term
is absent, {\it e.g.}, as explicitly verified for $SU(3)$ in (\ref{I_3}),
reflects the vanishing of the Euler characteristic of $SU(N)$ also for odd
$N$, as mentioned above. The in this case even dimension $N^2\!-\!1$ of
the algebra is irrelevant here, because, for the purpose of
cohomology, the parameter space of $SU(N)$ behaves as a product of
odd-dimensional spheres  $S^3\times S^5\times S^7 \times\cdots
S^{2N-1}$ \cite{Azcarraga1998}.   

The polynomials $I_N(\widehat m^2)$ do not have 
a constant term in either case and therefore vanish with $\widehat m^2\to
0$, {\it i.e.},  $I_N(0)=0$, as expected. Moreover, the scaling argument used
here and in the last section shows that the partition function
(\ref{Z_mGF}) of the massive Curci-Ferrari model can only depend on two of
the three parameters, 
\begin{equation} 
 Z_{\mbox{\tiny mGF}}(t,\xi,m^2) \,=\, f\big(t/\sqrt{\xi},\,\xi m^4\big) \; .
\label{Z_mGF-scale}
\end{equation}
An independent route of deriving this generic form, from the equations
of motions, will be presented below. In this section we explicitly 
obtained $f(0,y)$ with $y=\widehat m^4$ to constrain this function $f(x,y)$ of
two variables along the $x=t/\sqrt{\xi}=0$ line, and verified that 
\[
 Z_{\mbox{\tiny mGF}}(0,\xi,m^2) = f\big(0,\,\xi m^4\big) 
  \propto  \left\{ \begin{array}{ll} 
               (\xi m^4)^{\#  {\rm sites}/2} , & N=2 \\[4pt]
               (\xi m^4)^{\#  {\rm sites}} \, , & N=3 
                      \end{array} \right. 
\]
for $m^2 \to 0$. Because of the topological explanation of the zero
obtained in this limit, {\it i.e.}, $f(0,0)=0$, as discussed in the
last section, this actually constrains $f$ to vanish along the entire
$y=0$ line, $f(x,0)=0$ for all $x=t/\sqrt{\xi}$. 

For $x=0$ we can in principle therefore define a non-vanishing, finite
limit, 
\begin{equation} 
    \lim_{m^2\to 0} (\xi m^4)^{-N_\mathrm{tot}}  Z_{\mbox{\tiny
    mGF}}(0,\xi,m^2) \, = \, \mbox{const.} \label{naivelim}
\end{equation}
with an appropriate power $N_\mathrm{tot} = \# $ of sites on a 
finite lattice for odd $N$, or half that for even $N$. This constant 
could thus be inserted into the unfixed lattice gauge theory
measure without harm, {\it i.e.}, avoiding the zero in (\ref{Z-zero}). 
Because $x=t/\sqrt{\xi}=0$, however, this still has no effect
in terms of gauge-fixing by the Faddeev-Popov procedure either. We
need to get away from $x=0$, at least by a small amount, in order to suppress
those parts of the gauge orbits with large violations of the Lorenz
condition.  At a finite Curci-Ferrari mass $m^2$, however, this is
aggravated by the fact that the gauge-fixing partition function of the
Curci-Ferrari model is no-longer that of a topological model, and we
thus no-longer have the  $t$-independence (or $x$-independence) of
(\ref{t-indep}) either. We can therefore not as yet conclude at this point
that the constant in (\ref{naivelim}) will essentially remain
unchanged when going to some finite $x =t/\sqrt{\xi} \not=0$ as we
must.  

We are not quite there yet, and
we will therefore have to have a closer look at the parameter
dependence of the massive Curci-Ferrari model in the next section.

\section{Parameter Dependences}
\label{PD}

From Eqs.~(\ref{Z_mGF}) and (\ref{S_mGF}) or (\ref{mSGFtexpl}) we
immediately obtain the following (logarithmic) derivatives,
\begin{eqnarray} 
  t \smallfrac{\partial}{\partial
        t } \, \ln Z_{\mbox{\tiny mGF}}(t,\xi,m^2) &=& -i 
                   \big\langle\, (s \bar s - i m^2) \, t\,  V_U[g]\, 
          \big\rangle_{m^2} \, , \nonumber \\
  2\xi  \smallfrac{\partial}{\partial
        \xi } \, \ln Z_{\mbox{\tiny mGF}}(t,\xi,m^2) &=& \nonumber \\
             && \hskip -2cm -i 
                   \big\langle \, (s \bar s - i m^2) \big(-i\xi \sum_i \bar
        c^a_i c^a_i \big) \, \big\rangle_{m^2}
        \, , \nonumber \\
  m^2  \smallfrac{\partial}{\partial
        m^2} \, \ln Z_{\mbox{\tiny mGF}}(t,\xi,m^2) &=&  -
          \big\langle \, m^2 \, \mathcal{O}(t,\xi) \, \big\rangle_{m^2}  \; , 
\label{Z-derivs}
\end{eqnarray}
where the subscripts $m^2$ on the right denote expectation values within the
Curci-Ferrari model at finite mass. In particular, the derivative
w.r.t.~$m^2$ in the last line differs from (\ref{mderiv}) only in that
$m^2$ has not been set to zero here yet. All these expectation values
can, in general, depend on the link-field configuration $\{U\}$ which
acts as a background field to the model. Independence of $\{U\}$ is
only guaranteed to hold in the topological limit $m^2\to 0$.

From the definition of $\mathcal O$ in (\ref{LatKondOp}), we thus find that
\begin{eqnarray} 
\left(  t \smallfrac{\partial}{\partial
        t }   +   2\xi  \smallfrac{\partial}{\partial
        \xi }  -   m^2  \smallfrac{\partial}{\partial
        m^2}  \right) \, \ln Z_{\mbox{\tiny mGF}}(t,\xi,m^2) &=& \\
&& \hskip -1.6cm 
 - i \big\langle \, s\bar s \, \mathcal{O}(t,\xi) \, \big\rangle_{m^2}  \; .
\nonumber
\end{eqnarray}
The standard argument that the expectation value of an (anti-)BRST
exact operator vanishes does not hold at finite $m^2$. Neither are
BRST and anti-BRST variations nilpotent, nor is $\mathcal O$ invariant
under the BRST or anti-BRST transformations. However, the equations of
motion for (anti-)ghost and Nakanishi-Lautrup fields on the lattice,
{\it i.e.}, their lattice Dyson-Schwinger equations, can be used to
show that, indeed, 
\begin{equation}
\big\langle \, s\bar s \, \mathcal{O}(t,\xi) \, \big\rangle_{m^2}  = 0
\; ,
\end{equation}
even at finite $m^2$. Therefore,
\begin{equation}
\left(  t \smallfrac{\partial}{\partial
        t }   +   2\xi  \smallfrac{\partial}{\partial
        \xi }  -   m^2  \smallfrac{\partial}{\partial
        m^2}  \right) \, Z_{\mbox{\tiny mGF}}(t,\xi,m^2) \, =\, 0 \; .
\end{equation}
This differential equation entails that we can write the partition
function of the model in the generic form (\ref{Z_mGF-scale}).

As we already did in the previous sections, we therefore continue to
use the new parameters  $x=t/\sqrt{\xi} $ and $\widehat m^2 =
m^2\sqrt{\xi}\,$ from now on, writing
\begin{equation}
   Z_{\mbox{\tiny mGF}} \, \equiv \,    Z_{\mbox{\tiny
   mGF}}(x,\widehat m^2)  \; .
\end{equation}
Again using rescaled fields $\sqrt{\xi}\,  b \to b$, $\sqrt[4]{\xi}\,
\bar c \to \bar c$, $\sqrt[4]{\xi}\, c \to c$ and
with  $\sqrt[4]{\xi}\, \bar s \to \bar s$, $\sqrt[4]{\xi}\,
s \to s$, so that the (anti-)BRST transformations of Eq.~(\ref{BRSc}) --
(\ref{aBRSb}) remain formally unchanged, the only modification is the
replacement of  $m^2 $ by $\widehat m^2$ in those of the massive model
in Eqs.~(\ref{mBRSb}), (\ref{maBRSb}).  Correspondingly, all other
relations above are then converted by the formal replacements $\xi\to
1$, $t\to x$ and $m^2 \to \widehat m^2$. In particular,
\begin{eqnarray}
S_{\mbox{\tiny mGF}}(x) &=& i \,( s \bar s-i\widehat m^2) \, \Big( x\, 
 V_U[g] \, - \, \smallfrac{i}{2}  \sum_i \bar c_i^a c_i^a \Big) \\
   &=& \sum_i \,  \Big\{ - i x  \, b^a_i  F_i^a(U^g)
\, - {i} x \,\bar c^a_i  {M_{\mbox{\tiny FP}}}_i^a[c]\nonumber \\[-4pt]
&& \hskip 1cm +\, 
\smallfrac{1}{2} b_i^a b_i^a \, + \, \smallfrac{1}{8}\, 
(\bar c_i \times c_i)^2 \, \Big\} \, + \widehat m^2\, \mathcal{O}(x)
 \; , \nonumber 
\end{eqnarray}
with
\begin{equation}
\mathcal{O}(x)\, = \,  x\, 
 V_U[g] \, - \, {i}  \sum_i \bar c_i^a c_i^a  
\end{equation}
The two independent derivatives left, are readily  
read off 
in an analogous way to give 
\begin{eqnarray} 
   \smallfrac{\partial}{\partial
        x } \, \ln Z_{\mbox{\tiny mGF}}(x,\widehat m^2) &=& -i 
                   \big\langle\, (s \bar s - i \widehat m^2) \,  V_U[g]\, 
          \big\rangle_{\widehat m^2} \, , \nonumber \\
    \smallfrac{\partial}{\partial
       \widehat m^2} \, \ln Z_{\mbox{\tiny mGF}}(x,\widehat m^2) &=&  -
          \big\langle \, \mathcal{O}(x) \,
        \big\rangle_{\widehat m^2}  \; . 
\label{indep-Z-derivs}
\end{eqnarray}
In absence of a topological argument for the gauge parameter
independence at finite Curci-Ferrari mass, the best we can do to
achieve independence of $x=t/\sqrt{\xi}$ is to allow an $x$ dependent 
mass parameter $\widehat m^2 \equiv \widehat m^2(x)$. In particular, the 
$x=0$ results of the previous section are then to be interpreted as
being expressed in terms of $\widehat m^2(0)$. These results will
remain unchanged for $x\not= 0$, if we adjust the mass function
$\widehat m^2(x) $ with $x$ in the partition function $ Z_{\mbox{\tiny
   mGF}}$, accordingly. That is, if
\begin{eqnarray}
  0 &=&  \smallfrac{d}{dx} \,    Z_{\mbox{\tiny
   mGF}}\big(x,\widehat m^2(x)\big) \\ &=&
  \left( \smallfrac{\partial}{\partial x} + 
\smallfrac{d \widehat m^2}{dx} \, \smallfrac{\partial}{\partial
   \widehat m^2} 
 \right) \,    Z_{\mbox{\tiny mGF}}\big(x,\widehat m^2(x)\big) \; . \nonumber
\end{eqnarray}
From Eqs.~(\ref{indep-Z-derivs}) we see that this requires that
\begin{equation}
\smallfrac{d \widehat m^2}{dx} \, = \, -i \frac{   \big\langle\, (s
          \bar s - i \widehat m^2) \,  V_U[g]\ \big\rangle_{\widehat
          m^2}}{ \big\langle \, \mathcal{O}(x) \, \big\rangle_{\widehat m^2}} 
           \; .  \label{mhat-der}
\end{equation}
This might not appear to be a very profound insight, because we simply
arranged matters by hand to achieve gauge-parameter independence in
this way. The crucial question at this
point is, whether the tuning of the Curci-Ferrari mass parameter with
$x$ is possible independent of the link configuration $\{U\}$ which is
far from obvious here. Otherwise we would have to choose a different
trajectory in the parameter space $(x,\widehat m^2)$ for different
gauge orbits which would be of little use then, as far as the 
Faddeev-Popov gauge-fixing procedure is concerned. If it is possible,
on the other hand, we can then use the value of the mass $\widehat
m^2_0 = \widehat m^2(0)$ at $x=0$ to regulate the Neuberger zero and
use the $x$ and $\{ U\}$ independent, non-vanishing and finite constant
\begin{equation} 
    \lim_{\widehat m_0^2\to 0} (\widehat m_0^4)^{-N_\mathrm{tot}}
    Z_{\mbox{\tiny 
    mGF}}(x,\widehat m^2(x)) \, = \, \mbox{const.} \label{suitlim}
\end{equation}
%
%
as the starting definition of Faddeev-Popov gauge fixing on the lattice.
Then, of course, we would also expect that there should be a
topological meaning to this constant which is so far, however,
unfortunately unknown to us. 

\section{Orbit independent gauge-
parameter expansion of the CF mass} 

As we have seen in the previous section, the gauge-parameter independence of
the gauge-fixing partition function $ Z_{\mbox{\tiny mGF}}$ of the
massive Curci-Ferrari model will in general require the rescaled Curci-Ferrari
mass parameter $\widehat m^2 = m^2 \sqrt{\xi}$ to depend on the gauge
parameter in a non-trivial way, {\it i.e.}, $\widehat m^2 \equiv
\widehat m^2(x)$, via $x=t/\sqrt{\xi}$.  

Gauge-parameter independence requires the derivative of $\widehat
m^2(x)$ to be given by Eq.~(\ref{mhat-der}). Together with the
condition that $\widehat m^2(0) = \widehat m_0^2$ we can use this
equation to obtain the coefficients of $\widehat m^2(x)$ 
in a Taylor series expansion around $x=0$, where we can do explicit
calculations. Importantly, we can then verify that these coefficients
will not depend on the gauge orbit, {\it i.e.}, on (the class of
gauge-equivalent) link configurations $\{U\}$. Being based on the
tensor method of invariant integrations over the gauge group elements
per site $i$ of the lattice, this perturbative expansion of the
Curci-Ferrari mass parameter at small $x$ will in fact be 
gauge-orbit independent at every order. As always, of course, nothing can be
learned in such an expansion about possible non-analytic
contributions. We therefore assume the analyticity of the massive
Curci-Ferrari model in the `would-be-Gauss-Bonnet' limit $x\to
0$. This should surely be valid in the massless limit, but we need to assume
here that the limits $x\to 0$ and $\widehat m_0^2 \to 0$ can be 
interchanged, in addition. While this is all well on a finite lattice,
it certainly needs to be kept in mind when studying the model in the infinite
volume and continuum limits.

On a finite lattice, it is relatively straightforward to show
that $\widehat m^2$ is in fact independent of the gauge-parameter $x$ at
$1^\mathrm{st}$ oder in this expansion, {\it i.e.}, that
 \begin{equation}
      \smallfrac{d\widehat m^2}{dx}  \Big|_{x=0} \, = \,  0 \;
      . \label{zero_der} 
\end{equation}
This is simply because the numerator in Eq.~(\ref{mhat-der}) vanishes
at $x=0$ while the denominator is a finite number. To see this
explicitly, first consider at $x=0$   
Eq.~(\ref{qGaussBonnet}) with our new variables and rescaled fields, 
before the gauge-group integrations,
\begin{eqnarray}
 Z_{\mbox{\tiny
   mGF}}(0,\widehat m_0^2) &=& 
 \int d[g, b,\bar c, c]  \, \times \label{meas_0} 
\label{ZmGFx0} \\ && \hskip -2.5cm  
 \exp\Big\{- \sum_i \Big(\, \smallfrac{1}{2} b_i^a b_i^a \, + \,
   \smallfrac{1}{8}\,  (\bar c_i \times c_i)^2 \, - i \widehat m_0^2\,
   \bar c_i\! \cdot \! c_i \Big) \Big\}   \; . \nonumber  
\end{eqnarray}
As mentioned above, it decouples from the link-field configuration and
factorises. Relative to this partition function, we obtain,
\begin{equation} 
\langle \, i \bar c^a_i c_j^b\,\, \rangle_{\widehat m^2} \big|_{x=0} = \,
\frac{\delta^{ab} \, \delta_{ij}}{N^2-1} \,  \, \frac{I_N'(\widehat
  m_0^2)}{I_N(\widehat m_0^2)} \; , \label{aux_1}
\end{equation}
which is easily verified from the rules of Grassmann integration and
Eqs.~(\ref{qGaussBonnet})--(\ref{cbc-siteint}). Because $\mathcal O(0) = - i
\sum_i \bar c_i^a c_i^a $, the denominator in (\ref{mhat-der}) at
$x=0$ is obtained from the trace in (\ref{aux_1}),
\begin{eqnarray}
  \big\langle \, \mathcal{O}(x) \, \big\rangle_{\widehat m^2}  
\Big|_{x=0} \!\! &=&
 \big\langle -i \sum_i \bar c_i^a c_i^a \,\,\big\rangle_{\widehat m^2}
 \big|_{x=0}  \label{aux_2}  \\ 
&=& - (\# {\rm sites}) \times   \, \frac{I_N'(\widehat
  m_0^2)}{I_N(\widehat m_0^2)} \; ,\nonumber
\end{eqnarray}
which on a finite lattice with non-vanishing $\widehat m_0^2$ is finite.  

For the numerator in (\ref{mhat-der}) we have to compute
\begin{eqnarray}
   \smallfrac{\partial}{\partial
        x } \, \ln  Z_{\mbox{\tiny mGF}}(x,\widehat m^2) &=& -i 
                   \big\langle\, (s \bar s - i \widehat m^2) \,  V_U[g]\, 
          \big\rangle_{\widehat m^2} \, , \nonumber \\
 &&\hskip -2.8cm = \, 
  \big\langle\,  i \big(b,F\big) + i \big( \bar c , M_{\mbox{\tiny FP}}c\big) -
        \widehat m^2 V_U[g] \, \big\rangle_{\widehat m^2} \; ,
        \label{num_0} 
\end{eqnarray}
where the brackets are introduced for summation over site and color indices. 
At $x=0$, the $b$-field integration is Gaussian and the first term in
(\ref{num_0}), linear in $b$, therefore vanishes. Because the gauge
fields decouple from the measure in (\ref{meas_0}), Eq.~(\ref{aux_1}) 
produces the trace of the Faddeev-Popov matrix $M_{\mbox{\tiny
 FP}}$ in the second term on the right of (\ref{num_0}) for $x=0$.
With
\begin{equation} 
 \mbox{tr}\, M_{\mbox{\tiny FP}}(U^g) \, = \, - 2C_2^\rho    
\, V_U[g] \; , \label{trM}
\end{equation}
where $C_2^\rho $  is the quadratic Casimir invariant in 
representation $\rho$, $ C_2^f = (N\! - \! 1/N)/2$ in the fundamental one,
we see that the second becomes proportional to the third, and both
proportional to the expectation value of the Morse potential $V_U[g]$.
This is linear in $U^g$ and contains exactly one element $g_i$ (or
$g^\dagger_i$) in each and every term 
for which the invariant integration over the gauge group at the particular site
$i$ produces a zero. Therefore,
\begin{equation} 
  \big\langle \, V_U[g] \, \big\rangle_{\widehat m^2} \Big|_{x=0} = \,
  0 \; ,
\end{equation}
which establishes (\ref{zero_der}). It means that the derivative of
$\widehat m(x)$ is of the order $x$ near $x=0$, or
\begin{equation} 
  \widehat m^2(x) = \widehat m_0^2 \,  +  \, \mathcal O(x^2)
                 \; . \label{CFmasssmallx}
\end{equation}
In order to compute the constant in the second order term, and verify
that it is non-vanishing and independent of $\{ U\}$, we can consider
the derivative w.r.t.~$x$ of the numerator in (\ref{mhat-der}), or
\begin{eqnarray}
  \smallfrac{\partial^2}{\partial x^2} \, \ln Z_{\mbox{\tiny
                   mGF}}(x,\widehat m^2) &=&  \\
&&\hskip -3cm
  \big\langle\,  \Big(\,i \big(b,F\big) + i \big( \bar c ,
                   M_{\mbox{\tiny FP}}c\big) -  
        \widehat m^2 V_U[g] \,\Big)^2\, \big\rangle_{\widehat m^2}
                   \nonumber \\
  && \hskip -2cm
      - \,   
\big\langle\,  i \big(b,F\big) + i \big( \bar c , M_{\mbox{\tiny FP}}c\big) -
        \widehat m^2 V_U[g] \, \big\rangle_{\widehat m^2}^2 \; . \nonumber 
\end{eqnarray}
The second (disconnected) term again vanishes at $x=0$. Expanding the
square in the first, we once more use that every term linear in the
$b$-field will also vanish at $x=0$, and therefore,
\begin{eqnarray}
  \smallfrac{\partial^2}{\partial x^2} \, \ln Z_{\mbox{\tiny
                   mGF}}(x,\widehat m^2)\Big|_{x=0} &=& 
 \big\langle\,  \big(i b,F\big)^2  \big\rangle_{\widehat m^2\!,\,0} 
\label{aux_3} \\
&&\hskip -2cm
 +  \,  \big\langle\,  \big( i\bar c ,
                   M_{\mbox{\tiny FP}}c\big)^2  
             \big\rangle_{\widehat m^2\!,\,0} +\, \widehat m^4_0 \,
 \big\langle\,   V_U^2[g] \, \big\rangle_{\widehat m^2\!,\,0} 
                    \nonumber \\
  && \hskip -1cm
      - \, 2  \widehat m^2_0  \,   
\big\langle\, \big( i\bar c , M_{\mbox{\tiny FP}}c\big) \, 
      V_U[g] \, \big\rangle_{\widehat m^2\!,\,0}  
\; , \nonumber 
\end{eqnarray}
where we used shorthand $  \langle\dots \rangle_{\widehat
  m^2\!,\,0} \equiv  \langle\dots \rangle_{\widehat
  m^2}|_{x=0} $.

\begin{table*}
  \centering
\parbox{0.7\linewidth}{
  \begin{tabular}{l@{\hskip .2cm}|@{\hskip .4cm}>{}c@{\hskip
	.4cm}|@{\hskip .4cm}>{}c} 
    Expectation Value & $N=2$ & $N\ge 3$ \\[4pt]
\hline
&& \\[-2pt]
$\displaystyle  \big\langle\,  \big(i b,F\big)^2
         \big\rangle_{\widehat m^2\!,\,0} $ & 
                   $\displaystyle - 2D\,  (\#\mbox{\small sites})\;  3 $ &  
                            $\displaystyle - 2D\,  (\#\mbox{\small sites})\;
                                   \Big(N-\smallfrac{1}{N}\Big)  $  \\[6pt]
$\displaystyle  \big\langle\,  V_U^2[g]\, \big\rangle_{\widehat m^2\!,\,0} $ & 
                   $\displaystyle 2D\,  (\#\mbox{\small sites})\; 2  $ &  
                      $\displaystyle 2D\,  (\#\mbox{\small sites})  $  \\[4pt]
$\displaystyle  \big\langle\, \big(i\bar c, M_{\mbox{\tiny FP}}c\big)
         \,  V_U^2[g]\, \big\rangle_{\widehat m^2\!,\,0} $ &
           $\displaystyle -  2D\,  (\#\mbox{\small sites})\;
              \smallfrac{3+12 \widehat m_0^4 }{3 \widehat m_0^2 +4
                \widehat m_0^6 }   $ &   
                      $\displaystyle - 2D\, (\#{\rm sites}) \;
                   \smallfrac{1}{N} \,
                   \smallfrac{I_N'(\widehat m_0^2)}{I_N(\widehat 
                                                           m_0^2)}  $  \\[6pt]
$\displaystyle  \big\langle\,  \big( i\bar c ,  M_{\mbox{\tiny
               FP}}c\big)^2  \big\rangle_{\widehat m^2\!,\,0} $ & 
               $\displaystyle -  2D\,  (\#\mbox{\small sites})\;
              \smallfrac{6 \widehat m_0^2 }{3 \widehat m_0^2 +4
                \widehat m_0^6 }   $ &   
                      $\displaystyle - 2D\, (\#{\rm sites}) \;
                   \smallfrac{2}{N} \,
                   \smallfrac{{J_N}(\widehat m_0^2)}{I_N(\widehat 
                                                           m_0^2)}  $  \\[6pt]
$\displaystyle  \big\langle \, \mathcal{O}(x) \, \big\rangle_{\widehat
             m^2\!,\,0} $ &  
           $\displaystyle -  (\#\mbox{\small sites})\;
              \smallfrac{3+12 \widehat m_0^4 }{3 \widehat m_0^2 +4
                \widehat m_0^6 }   $ &   
                      $\displaystyle - (\#{\rm sites}) \;
                      \smallfrac{I_N'(\widehat m_0^2)}{I_N(\widehat
                                                           m_0^2)}  $  
  \end{tabular}
  \caption{The individual terms in Eq.~(\ref{aux_3}) for the numerator of
    the derivative in (\ref{mhat-der}) in the limit $x\to 0$ together with the
    leading contribution to the denominator from Eq.~(\ref{aux_2}).} }
   \addtocounter{table}{-1}\refstepcounter{table}\label{res_tab}
\end{table*}

We calculate and discuss each of the individual terms on the right of
(\ref{aux_3}) separately in Appendix~\ref{expl_calcs} (with the
exception that, by the same argument that led to (\ref{trM}), the last
two are essentially the same again, {\it i.e.}, both $\propto \langle\,
V_U^2\, \rangle_{\widehat  m^2\!,\,0} $). In particular, 
we show there that they are indeed all independent of $\{U\}$. The
results for $SU(N)$ gauge-groups in $D$ Euclidean dimensions 
are summarized in Table~\ref{res_tab}, where $I_N$ is
defined in Eq.~(\ref{cbc-siteint}) and $J_N$ analogously by
\begin{eqnarray} 
 J_N(\widehat m^2) & = & \int \prod_{a=1}^{N^2\!-1} d(i\bar c^a) d c^a
\, \left(  -\smallfrac{1}{8} (\bar c \times c)^2 \right)
 \, \label{cbc-int-siteint}\\
&& \hskip 1.4cm
 \exp\left\{ i \widehat m^2\,  \bar c\! \cdot \! c \, - \,
 \smallfrac{1}{8}\,   (\bar c \times c)^2 \right\} \; . \nonumber
\end{eqnarray}
A comparison of the integral expression for $J_N$ with the analogous one for
  $I_N$ in Eq.~(\ref{cbc-siteint})
shows that with an explicit interaction inserted in the
integral, $J_N(\widehat m^2)$ is a polynomial in $\widehat m^2$ of two
orders less than $I_N(\widehat m^2)$. Just as $I_N(\widehat m^2)$, this
polynomial has no zero-order, constant term because the Euler
characteristic of $SU(N)$ vanishes and we essentially obtain a Gauss-Bonnet
integral for $\widehat m^2 \to 0$ again. Therefore,  
\begin{equation}
  \label{aux41}
  J_N(\widehat m^2) \sim I_N(\widehat m^2)\, \to 0 \; , \;\; \mbox{for }
  \widehat m^2 \to 0\;.
\end{equation}
Explicitly, for $SU(2)$, see Appendix \ref{expl_calcs},
\begin{equation}
  \label{J_2}
  J_2(\widehat m^2) \, =\, \smallfrac{1}{8} \, I_2''(\widehat m^2) \, =\,
  \smallfrac{3}{4} \, \widehat m^2 \; , 
\end{equation}
while for $SU(3)$ a straightforward but a bit more tedious computation
analogous to that used for Eq.~(\ref{I_3}) yields,
\begin{equation}
  \label{J_3}
  J_3(\widehat m^2) \, =\, \smallfrac{135}{64} \, \widehat m^4 \,+\, 
\smallfrac{45}{8}\,  \widehat m^8\, + \, 3\, \widehat m^{12} \; . 
\end{equation}

With the results in Table~\ref{res_tab} we can go back to the 
derivative of the Curci-Ferrari mass parameter $\widehat m^2(x)$ 
w.r.t.~$x$. Recall that the independence of the massive
Curci-Ferrari model on the gauge parameter $x=t/\sqrt{\xi}$ requires 
this derivative to be given by the ratio, {\it c.f.},
Eqs.~(\ref{mhat-der}) and (\ref{indep-Z-derivs}),   
\begin{equation}
\smallfrac{d \widehat m^2}{dx} \, = \, \frac{   \smallfrac{\partial}{\partial
        x } \, \ln Z_{\mbox{\tiny mGF}}(x,\widehat m^2) }{ \big\langle
        \, \mathcal{O}(x) \, \big\rangle_{\widehat m^2}}  
           \; .  \label{app_mhat-der}
\end{equation}
With (\ref{aux_3}) and the results from the table, in the limit
 $x\to 0$ we therefore find,
\begin{eqnarray}
\smallfrac{d \widehat m^2}{dx} &\stackrel{x \to 0}{\longrightarrow} & 
x\, 2D\, \Big\{ \, \smallfrac{N^2-1}{N}\, 
\smallfrac{I_N(\widehat m_0^2)}{I_N'(\widehat m_0^2)} \label{app_c2} \\
&&  \, -\, \smallfrac{2}{N} \, \smallfrac{{J_N}(\widehat
  m_0^2)}{I_N'(\widehat m_0^2)} \,- \, \widehat m_0^4 \,
\smallfrac{I_N(\widehat m_0^2)}{I_N'(\widehat m_0^2)} \, -\, 2 \widehat
m_0^2 \, \smallfrac{1}{N} \Big\} \; ,
\nonumber
\end{eqnarray}
for $N\ge 3$, and
\begin{eqnarray}
\smallfrac{d \widehat m^2}{dx} &\stackrel{x \to 0}{\longrightarrow} & 
x\, 2D\, \Big\{ \, 3 \,
\smallfrac{I_2(\widehat m_0^2)}{I_2'(\widehat m_0^2)} 
 \, -\, \smallfrac{1}{4} \, \smallfrac{I_2''(\widehat
  m_0^2)}{I_2'(\widehat m_0^2)}  \\
&&  \hskip -.6cm - \, 2\widehat m_0^4 \,
\smallfrac{I_2(\widehat m_0^2)}{I_2'(\widehat m_0^2)} \, -\, 2 \widehat
m_0^2 \Big\} \, 
= \, - x\, 2D\, \widehat m_0^2 \, \gamma_2(\widehat
m_0^2) \; ,
\nonumber
\end{eqnarray}
for $SU(2)$, where, using Eq.~(\ref{I_2}) for $I_2$, 
\begin{equation}
\gamma_2(\widehat m_0^2) \, = \,  \frac{3 + 18\, \widehat m_0^4 +
  8\, \widehat m_0^8}{3 + 12 \,  \widehat m_0^4}\; . \label{gamma_2}
\end{equation}
In either case, the expansion of the Curci-Ferrari mass parameter
around $x=0$ for $SU(N)$ in $D$ dimensions can finally 
be written in the general form
\begin{equation} 
        \widehat m^2(x) \, =\, \widehat m_0^2 \,  - \, x^2\, D\, \widehat
        m_0^2 \, \gamma_N(\widehat m_0^2)      \, + \, \mathcal
        O(x^3) \; ,
\end{equation}
where $\gamma_N(\widehat m_0^2) $ is a ratio of polynomials in
$\widehat m_0^2 $ obtained via (\ref{app_c2}), for $N\ge 3$, from
\begin{eqnarray} 
  \gamma_N(\widehat m_0^2) &=& 
   \smallfrac{2}{N}\,  \left(1+ \smallfrac{{J_N}(\widehat
  m_0^2)}{\widehat m_0^2 I_N'(\widehat m_0^2)}\right) \label{app_c6} \\
&& \hskip 1cm - \,
\left(\smallfrac{N^2-1}{N} - \widehat m_0^4 \right) \,
\smallfrac{I_N(\widehat m_0^2)}{\widehat m_0^2 
 I_N'(\widehat m_0^2)} \; .\nonumber
\end{eqnarray}
For $N = 3$, using (\ref{I_3}) and (\ref{J_3}), we find explicitly,
\begin{eqnarray}
\gamma_3(\widehat m_0^2) &=& \frac{1}{3} \, \times
\label{gamma_3} \\[4pt]
 &&\hskip -.6cm    \frac{90 + 855\,
  \widehat m_0^4 +  1692\, \widehat m_0^8 + 1088\, \widehat m_0^{12} + 192
  \, \widehat m_0^{16}}{90 + 720 \, \widehat m_0^4 + 1152 \, \widehat
 m_0^8 + 512  \,  \widehat m_0^{12}} \; . \nonumber
\end{eqnarray}
To work out $\gamma_N$ for general $N$ in the limit $\widehat m_0^2\to
0$, remember that $I_N(\widehat m_0^2)$ is an odd/even polynomial in
$\widehat m_0^2 $ for $N$ even/odd. In either case there is no
constant term, $I_N(\widehat m_0^2)\to 0$ for  $\widehat m_0^2\to
0$. Therefore
\begin{equation} 
\frac{I_N(\widehat m_0^2)}{\widehat m_0^2 I_N'(\widehat m_0^2)} \,
\stackrel{\widehat m_0^2 \to 0}{\longrightarrow} \, \left\{ 
\begin{array}{cl}
1 \, ,& N\;\; \mbox{even,} \\[4pt]
\smallfrac{1}{2} \, ,& N\;\; \mbox{odd.} 
\end{array} \right.  \label{app_c8}
\end{equation}
For the leading power $n$ of $\widehat m^2_0$ 
in $I_N(\widehat m_0^2)$ near $\widehat m^2_0=0$, which is $n=1$ when $N$ is
even and $n=2$ when $N$ is odd, we need to expand the exponential of
the ghost self-interactions in its integral representation
(\ref{cbc-siteint}) to a power $p$ such that  
\[ n \, + \, 2p =  N^2 - 1 \; . \] 
Otherwise the Grassmann integrations over $N^2\!-\!1$ ghost (and
anti-ghost) variables will produce zero. Therefore,
\begin{equation} 
  p \, = \, \left\{ 
\begin{array}{cl}
 \smallfrac{N^2-2}{2} \, ,& N\;\; \mbox{even,} \\[6pt]
\smallfrac{N^2-3}{2} \, ,& N\;\; \mbox{odd.} 
\end{array} \right. \label{app_c9}
\end{equation}
Comparing the definition of $J_N$ in Eq.~(\ref{int-siteint}) to that of
$I_N$ in (\ref{cbc-siteint}), we see that the exponential of
the ghost self-interactions in the integral representation of $J_N$
needs to be expanded to one power less, {\it i.e.}, to the power $p-1$
for the leading term. Comparing the prefactors of these terms in
each case we therefore find,
\begin{equation}
\frac{J_N(\widehat m_0^2)}{I_N(\widehat m_0^2)} \,
\stackrel{\widehat m_0^2 \to 0}{\longrightarrow} \,
\smallfrac{p!}{(p-1)!} \, =\,  p\; .
 \end{equation}
With (\ref{app_c8}) and (\ref{app_c9}) this then implies,
 \begin{equation} 
\frac{J_N(\widehat m_0^2)}{\widehat m_0^2 I_N'(\widehat m_0^2)} \,
\stackrel{\widehat m_0^2 \to 0}{\longrightarrow} \, \left\{ 
\begin{array}{cl}
 \smallfrac{N^2-2}{2} \, ,& N\;\; \mbox{even,} \\[6pt]
\smallfrac{N^2-3}{4} \, ,& N\;\; \mbox{odd.} 
\end{array} \right.  \label{app_c11}
\end{equation}
For even $N \ge 4$ we thus obtain from Eq.~(\ref{app_c6}),
\begin{equation}
  \label{app_c12}
  \gamma_N \,\to \, \smallfrac{2}{N} \,
  \Big(1+\smallfrac{N^2-2}{2}\Big) - \smallfrac{N^2-1}{N} \, =\,
  \smallfrac{1}{N} \; ,   
\end{equation}
and twice that for $SU(2)$, {\it c.f.}, Eq.~(\ref{gamma_2}), where 
$\gamma_2 \to 1 = 2/N$ for $N=2$, in the limit $\widehat m_0^2\to 0$. 
This doubling in the $SU(2)$ case essentially comes about because for
the expectation values containing terms which are at most quadratic in
the link variables $ U$ at this order, only for $N=2$ we obtain
contributions from two types of invariant integrals, Eqs.~(\ref{invint_1}) and
(\ref{invint_2}), while only the gauge group integrations of the form in
(\ref{invint_1}) contribute for $N\ge 3$ at this order (additional
contributions similar to those for $SU(2)$ here, will arise, {\it
  e.g.} for $SU(3)$ at the next order etc.).

For all odd $N\ge 3$, at the present
order, (\ref{app_c8}) and (\ref{app_c9}) therefore give,
\begin{equation}
  \label{app_c13}
  \gamma_N \,\to \, \smallfrac{2}{N} \,
  \Big(1+\smallfrac{N^2-3}{4}\Big) - \smallfrac{N^2-1}{2N} \, =\,
  \smallfrac{1}{N} \; ,   
\end{equation}
which again leads to the same result as obtained for the even $N \ge
4$ above, {\it i.e.},
\begin{equation} 
  \label{app_c14}
 \gamma_N(\widehat m_0^2) \,\to \, \smallfrac{1}{N} \; ,
 \;\;\mbox{for}\;\ \widehat m_0^2 \,\to\,
0 \;, 
\end{equation}
and all $N\ge 3$. 

All these results are gauge-orbit independent as they must.
While this is merely necessary, but not
sufficient, it demonstrates that we can get away from $x=0$, at least
perturbatively in a small $x$ expansion. This is of qualitative
importance as a non-zero value of $x=t/\sqrt{\xi}$, no matter how
small, corresponds to a large but finite $\xi$ at $t=1$ and thus
eliminates the gauge freedom.

In summary, the gauge-orbit independence of the constant in the second
order term of the small $x$ expansion  of $\widehat m^2(x)$ is a
direct consequence of the invariant integrations over the gauge-group
elements at each site. The gauge-group integrations can in fact be
performed at any order in this Taylor expansion around $x=0$ because
the action is independent of the gauge group there ({\it c.f.},
Eq.~(\ref{ZmGFx0})).  
Moreover, the invariant tensor method can be used to demonstrate how 
these integrations will ensure that the coefficients in this Taylor
expansion are indeed independent of $\{U\}$ at any order in $x$. This
gauge-orbit independence of the mass expansion is verified explicitly
for the constant in the second order term of (\ref{CFmasssmallx}) from
the results in Appendix~\ref{expl_calcs}. In particular, as we have
shown above, in $D$ Euclidean dimensions the gauge-parameter
expansion of the Curci-Ferrari mass in Eq.~(\ref{CFmasssmallx}) becomes,
\begin{equation} 
  \widehat m^2(x) = \widehat m_0^2 \, \Big( 1 - D \,  \gamma_{N}(\widehat
  m_0^2)\, x^2  + \mathcal O(x^3) \Big) \; ,
  \label{CFmasssmallxm} 
\end{equation}
where $\gamma_2$ and $\gamma_3 $ for $SU(2)$ and $SU(3)$ are
explicitly given by Eqs.~(\ref{gamma_2}) and (\ref{gamma_3}).
Moreover, at leading order in the Curci-Ferrari mass parameter, 
$\gamma_N(\widehat m_0^2) $ is finite and of the order $1/N$ in the
limit $\widehat m_0^2 \to 0$.  
For general $SU(N)$ gauge groups we found,
\begin{equation} 
        \gamma_N(\widehat m_0^2) = \left\{\begin{array}{ll} 
             1 \, +\, \mathcal O(\widehat m_0^4) \, , & N=2 \, ,\\[4pt]
            \frac{\displaystyle 1}{\displaystyle N}\, + \, \mathcal
        O(\widehat m_0^4) \, , & N\ge 3 \, . 
					 \end{array}\right.
\end{equation}
The leading $x$-dependence of the Curci-Ferrari mass is therefore 
order $N$ suppressed. 
Without need to adjust the Curci-Ferrari mass parameter $\widehat m^2$
with $x$, the gauge-fixing partition function $Z_{\mbox{\tiny mGF}}$ of the
massive Curci-Ferrari model therefore becomes gauge parameter
independent at least up to the order $x^3$ in the large $N$ limit.

\section{Summary} 
\label{Summary}

We have formulated the Curci-Ferrari model on the lattice.
In the massless case this model provides an explicit demonstration of
the topological origin of the Neuberger 0/0 problem of lattice BRST.
The starting point of Neuberger's original argument was the observation of 
uncompensated Grassmann integrations producing a zero result in a
certain limit. This same limit in the massless Curci-Ferrari model
with its double BRST symmetry and quartic ghost self-interactions
corresponds to the Gauss-Bonnet limit, $ \xi \to \infty$, of a
topological model that computes the Euler characteristic of the
lattice gauge group which vanishes for compact Lie groups.  The fact
that the Neuberger zero is independent of this special limit then
follows directly from the gauge-parameter $\xi $ independence of the
topological model.

Introducing a Curci-Ferrari mass term regularizes the Neuberger zero.
The analogue of the Gauss-Bonnet-Neuberger limit here corresponds to
the gauge-parameter $\xi\! \to\! \infty$ limit together with the
Curci-Ferrari mass $m^2\to 0 $ such that the product $m^2\sqrt{\xi} $
remains finite, {\it i.e.}, $m^2\sqrt{\xi} \to \widehat m^2$ for some
finite $\widehat m^2$. In this limit, the partition
function of the massive Curci-Ferrari model on a finite lattice is
obtained as a polynomial in the new mass parameter $\widehat m^2$ and
is hence non-vanishing. The $0/0$ problem is thus avoided. 
However, the massive Curci-Ferrari model is no-longer
a purely topological model. BRST and anti-BRST are explicitly broken
by the Curci-Ferrari mass. The $sl(2, \RR ) \rtimes $ double BRST
algebra of the massless Curci-Ferrari model is decontracted into a simple
superalgebra for $m^2 \not=0$. As a result of this BRST breaking,
meaning that neither BRST nor anti-BRST transformations are nilpotent
anymore, the gauge-fixing partition function of the massive
Curci-Ferrari model is a priori not independent of the gauge parameter
$\xi $. This implies that the Curci-Ferrari mass has to depend on $\xi $  so as
to restore total $\xi$-independence, a requirement which in turn 
allows to determine this $\xi$-dependence of $m^2$. A gauge-orbit independent 
 Taylor series expansion of $m^2(\xi) $, in $D$ dimensions of the
 form, {\it c.f.}, Eq.~(\ref{CFmasssmallxm}), 
\begin{equation} 
  m^2(\xi) \, = \, \smallfrac{\widehat m_0^2}{\sqrt{\xi}} \, \Big( 1 -
  \smallfrac{D \gamma_N}{\xi} + \cdots \Big) \; ,
\end{equation}
around the Gauss-Bonnet limit of $1/\xi
\to 0$ is possible with $\xi$-independent mass parameter $\widehat
 m_0^2$, to show that one can meaningfully define a
   limit $\widehat m^2_0 \to 0$ in the spirit of l'Hospital's rule.
 In this limit, $\gamma_N \to 1/N$, for $SU(N)$ gauge theory with
   $N\ge 3$ (and $\gamma_2 \to 1$ for $SU(2)$).

\section{Conclusions and Outlook}
\label{CaO}

The explicit BRST breaking by the Curci-Ferrari mass term is
well-known to lead to unitarity violations in the continuum quantum
field theory. The BRST-cohomology construction of a positive physical
Hilbert space breaks down together with this BRST breaking. 
There are explicit examples of negative norm states
mixing into what would otherwise be defined as the physical subspace
\cite{Ojima1982,deBoer1996}.  For that reason, the parameter $m^2$
should not be interpreted as a physical mass but rather only as a
regulator for the Neuberger 0/0 problem of lattice BRST. 

In the Landau gauge limit with $\xi=0$, for example, it has the effect
of reweighting different Gribov copies depending on their value for the
Morse potential on the gauge orbit, $V_U[g]$,  which avoids their perfect
cancellation. The analogue of the
Poincar\'e-Hopf theorem for the Euler characteristic of the lattice
gauge group, Eq.~(\ref{P-H}) for finite $m^2$, becomes, 
\begin{eqnarray} 
  Z_{\mbox{\tiny mGF}}(t,0,m^2) 
 &=& \label{P-H-ext} \\[2pt]
&&\hskip -2.5cm  \sum_{\mbox{\tiny copies $\{g^{(i)}\!\}$}} \hskip -.2cm
  \mbox{sign} \,\big( \mbox{det} \, M_{\mbox{\tiny FP}}(U^{g^{(i)}})
  \big) \; \exp\big\{ - m^2 t\,  V_U[g^{(i)}] \big\} 
\; . \nonumber
\end{eqnarray}
At finite $m^2$ this leads to a suppression of Gribov copies outside
the fundamental modular region, {\it i.e.}, a suppression of all
copies relative to the absolute minima of $V_U[g]$.

In particular, because we furthermore have 
$V_U[g] \propto -  \mbox{tr}\, M_{\mbox{\tiny  FP}}$, 
each positive (negative) eigenvalue will increase (suppress) the
weight of a given copy. The situation is thus similar to that in the
Gribov-Zwanziger approach \cite{Gribov1978,Zwanziger1993} which
includes a horizon functional to suppress Gribov copies with negative
eigenvalues, outside the first Gribov region. And as in the massive
Curci-Ferrari model, this leads to a certain BRST breaking in the
Gribov-Zwanziger framework also. While the renormalizability of this
framework is maintained \cite{Zwanziger1993}, 
the unitarity violations when attempting a
BRST cohomology construction of a physical Hilbert space remain to be
a problem there as well. In fact, the horizon condition might be
interpreted in a way as to generate a kind of Curci-Ferrari mass
dynamically. 

The so-called soft BRST breaking of the Gribov-Zwan\-zi\-ger framework, 
which really means that it is a non-per\-turba\-tive BRST breaking,  
should not be a problem, if it is effective in restricting the Landau
gauge to the fundamental modular region, the set of absolute minima
along the gauge orbits \cite{Maggiore1993}. This should be intuitively
expected, of course, for a perfect gauge-fixing, but because of the
existence of many degenerate absolute minima it is not mathematically
obvious. Note that we could at least formally achieve the same
restriction here, if we define the Landau gauge limit of the massive
Curci-Ferrari model as the limit of vanishing gauge-parameter, $\xi
\to 0$, at  finite mass parameter $\widehat m^2 = m^2 \sqrt{\xi}$
(analogous to the opposite limit $\xi\to\infty$ at fixed $\widehat
m^2$ considered above), 
as this would imply $m^2 \!\to \!\infty $ in (\ref{P-H-ext}), so as to
suppress all copies but the absolute minima of $V_U[g]$.

The arguments for the Gribov-Zwanziger framework  to achieve this
further restriction to the absolute minima in lattice formulations
\cite{Zwanziger1992}, furthermore involve the thermodynamic or
infinite volume limit, in which the common boundary of the fundamental 
modular region and the first Gribov region dominate the minimal Landau
gauge configuration space \cite{Zwanziger2004}.  To find absolute
minima in lattice simulations is not feasible for large lattices as this
is a non-polynomially hard computational problem. When sampling the
local minima of the first Gribov region on a finite lattice, as is
usually done in minimal lattice Landau gauge implementations
numerically, BRST breaking and unitarity violations could therefore be
a potential problem there also.  

In fact, related to this potential problem of minimal Landau gauge
implementations on finite lattices might be the question of a gluon
mass and the infrared behavior of the gluon and ghost propagators in
Landau gauge. While early lattice studies of those propagators 
\cite{Leinweber:1998im,Bloch:2003sk}
supported the predicted infrared behavior based on their
Dyson-Schwinger equations qualitatively well
\cite{vonSmekal:1997is,vonSmekal:1997vx}, small but significant
differences are increasingly being observed nowadays
in detailed comparisons and studies of finite-volume effects
\cite{Fischer:2007pf,Fischer:2007mc,Sternbeck:2007ug,Cucchieri:2007md}.
In particular, the Kugo-Ojima confinement criterion of local quantum
field theory seems now violated in the infinite-volume limit:  

Based upon the assumption that BRST-invariance is unbroken, the
continuum prediction is that the infrared dominant correlations are
mediated by the Faddeev-Popov ghosts, whose propagator is infrared
enhanced in Landau gauge, while the gluon propagator is found to be
infrared suppressed. This infrared behavior is now completely
understood in terms of confinement in QCD
\cite{Alkofer2001,Alkofer:2000mz}, it is a
consequence of the celebrated Kugo-Ojima confinement criterion. The
subsequent verification of this infrared behavior with a variety of
different functional methods in the continuum meant a remarkable
success
\cite{Lerche:2002ep,Zwanziger:2001kw,Pawlowski:2003hq}.  
In fact, it is directly tied to the validity and
applicability of the framework of local quantum field theory for
non-Abelian gauge theories beyond perturbation theory. 
Consistent with the conditions for confinement in local quantum field
theory, these predictions all lead to a conformal infrared behavior
for gluonic Green's functions which is yet to be observed in lattice
simulations. Because simulations must necessarily be done in a finite
volume, where such a behavior can strictly never be observed,
finite-volume effects have long been blamed for the remaining mismatch
with the continuum studies. Recently, these effects have been analyzed
carefully in the Dyson-Schwinger equations
\cite{Fischer:2007pf,Fischer:2007mc}. These results together with  
latest lattice data  on huge
lattices \cite{Sternbeck:2007ug,Cucchieri:2007md}, corresponding
to physical lengths of more than 20 fm in each direction, clearly rule
out finite-volume effects as the sole origin for the observed
discrepancies.     

Rather, the most likely origin for these discrepancies potentially hints at a
much more profound problem: a BRST breaking that a sampling of minima of
the gauge-fixing potential in lattice simulations might bring about
much like the restriction to the first Gribov region does in the
Gribov-Zwanziger framework. The observed evidence of a massive
infrared behavior of the gluon propagator in the infinite volume
limit in current lattice Landau-gauge implementations in fact suggests
that. Any reweighting of Gribov copies, inside or outside the 
first Gribov region, should correspond to a BRST breaking procedure
analogous to the introduction of a Curci-Ferrari mass. Only if the
numerical procedure converges towards a correct sampling of the
fundamental modular region in the infinite volume and continuum limits,
can the BRST breaking effects be expected to go away, which they will then
probably do together with the observed gluon mass. 


In interesting alternative procedure based on 
stereographic projection to define lattice gauge fields is
provided by the modified lattice Landau gauge of
Ref.~\cite{vonSmekal:2007ns}. This gives rise to a manifestly BRST
invariant lattice formulation. The Neuberger 0/0 problem is avoided
there because it is not the vanishing Euler characteristic of the lattice gauge
group that is calculated by the gauge-fixing partition function in
this case, but that of a stereographically projected manifold. In this
approach the pure lattice-artefact Gribov problem of compact
$U(1)$ is avoided because the Faddeev-Popov operator is positive, and
there are thus no cancellations between Gribov copies 
\cite{vonSmekal:2007ns}. Consequently, there are none along the maximal
Abelian subgroup $U(1)^{N-1}$ of $SU(N)$ either. This is just enough, however,
to remove the complete cancellation of Gribov copies in $SU(N)$
also. The remaining cancellations between copies of either
sign in $SU(N)$, which persist in the continuum limit, are necessarily
incomplete because the Euler characteristic of the coset manifold is
non-zero. It is essentially determined by that of the even dimensional spheres
$S^2\times S^4\times\cdots S^{2N-2} $ or, more precisely, of the
corresponding even-dimensional, real projective spaces 
$\mathbb{R}\mathrm{P}^{2n}$, of one dimension less than the odd-dimensional
spheres of the group manifold. 
The perhaps most promising feature of the modified lattice Landau
gauge is, however, that it provides a way to perform gauge-fixed
Monte-Carlo simulations sampling all Gribov copies of either sign in
the spirit of BRST. Bridging the gap between gauge-fixed Lattice and
continuum studies this should, in particular, resolve the present
discrepancies observed in the QCD Green's functions once and for all,
and at the same time put our theoretical knowledge of QCD and all
gauge theories of the Standard Model on solid ground in a completely
non-perturbative manner.

Meanwhile, the decontracted double BRST supersymmetry of the massive
Curci-Ferrari model on the lattice provides an interesting testbed
with a controlled BRST breaking and regularized Neuberger zeroes. It
might have its own interesting topological features and interpretation
as indicated by its potential gauge-orbit and gauge-parameter independence. 
This certainly deserves further study, especially with regard to its
potential to generalize the topological field theory relation between
the Gauss-Bonnet and Poincar\'e-Hopf theorems
\cite{Birmingham1991,AlvarezGaume:1983at,Mathai:1986tc}. In fact,  an
extension to Morse theory similar to (\ref{P-H-ext}), but corresponding to
the introduction of an imaginary Curci-Ferrari mass parameter ({\it
  i.e.}, using a real parameter $\phi\equiv im^2t$ in
(\ref{P-H-ext})), and its relation to a generalized Gauss-Bonnet
theorem reminiscent of (\ref{qGaussBonnet}) were discussed in
Ref.~\cite{Niemi:1994ej}. This might help to understand how it is
possible to achieve the gauge-parameter and orbit independence in the
massive Curci-Ferrari model in general.  And it might 
provide an interesting new topological interpretation of the
Curci-Ferrari model within the decontracted double BRST $osp(1|2)$
superalgebra framework. 



\acknowledgments
\noindent
This research was supported by the Australian Research Council.

\appendix

\section{BRST Derivation of Faddeev-Popov Operator and Gauge-Fixing Action}
\label{expl_ders}

In this appendix we show explicitly how the double (anti-)BRST
variation in the gauge-fixing action of the topological model in
Eq.~(\ref{S_GF}) leads to the explicit form given in (\ref{SGFexpl}).
This lattice transcription of a well-known continuum procedure is
mainly given for the reader's convenience and ready reference. 
 
Performing the anti-BRST variation on the r.h.s.~in Eq.(\ref{S_GF})
first, we obtain 
\begin{eqnarray}
  \bar s \, V_U[g] &=& 
 \smallfrac{1}{2\rho} \, \sum_i \sum_{j\sim i} \mbox{tr}\,\big(
 \bar c_i  U^g_{ij} - \bar c_j  U^g_{ij} \big) \\
 &=&  -  \sum_i  \bar c_i^a  F_i^a(U^g) \; , \;\; \mbox{where} \nonumber\\
 F^a_i(U^g) &=&  -\smallfrac{1}{2\rho} \,  \sum_{j\sim i}
 \mbox{tr}\,\big( X^a(  U^g_{ij} -   U^g_{ji}) \big) 
\label{std-gfc_app}
\end{eqnarray}
is used in the standard gauge-fixing condition of covariant gauges.
In the continuum limit it reduces to
\begin{equation} 
 F^a_i(U^g) \, \stackrel{a\to 0}{\longrightarrow} 
 \,  a^2\, \partial_\mu {A_\mu^g}^a  \, +\, \mathcal{O}(a^4) \; .
\end{equation}
With Eqs.~(\ref{aBRSc}), (\ref{aBRScb}) we furthermore have
\begin{equation}
\bar s \, \big( \bar c^a c^a \big) \, = \, \bar c^a b^a \; ,
\label{sbcbc}
\end{equation}
and therefore, for the gauge-fixing action, the alternative form
\begin{equation}
S_{\mbox{\tiny GF}} \, = \, -i \, \sum_i \,
s  \Big(  \bar c_i^a  \, \big( F_i^a(U^g) \, + \smallfrac{i\xi}{2} \, 
  b_i^a \big)  \Big) \; . \label{S_GF-standard}
\end{equation}
As in the continuum formulation, in this form it looks exactly like
the gauge-fixing action of standard Faddeev-Popov theory for the
linear covariant gauge. The specific features of the ghost/anti-ghost
symmetric framework show when working out the remaining BRST
variation. From the first term we have ({\it i}),
\begin{eqnarray}
\sum_i \big( s \bar c_i^a \big) \, F_i^a  &=&  
-  \smallfrac{1}{2\rho} \, \sum_i \sum_{j\sim i}\, \mbox{tr}\,\big(   
 b_i  (  U^g_{ij} -   U^g_{ji}) \big) \label{var1} \\
 && \; +\,
\smallfrac{1}{4\rho} \, \sum_i \sum_{j\sim i}\, \mbox{tr}\,\big(   
 \{ \bar c_i,\, c_i\}\,  (  U^g_{ij} -   U^g_{ji}) \big) \;. \nonumber
\end{eqnarray}
Herein, the first term on the right implements the gauge-fixing
condition as in standard Faddeev-Popov theory. 
The second term, containing the anticommutator $
\{ \bar c,\, c\}$, is characteristic of ghost/anti-ghost symmetry
because it combines with the remaining quadratic ghost terms to produce
a Hermitian Faddeev-Popov operator (for any gauge parameter $\xi$). To
see this explicitly, consider ({\it ii}),
\begin{eqnarray}
\sum_i \bar c_i^a  \, \big( s F_i^a )  &=&  
\smallfrac{1}{2\rho} \, \sum_i \sum_{j\sim i}\, \mbox{tr}\,\big(   
 \bar c_i c_i  U^g_{ij} \label{stdFP}\\
 && \;\;   -   \bar c_i  U^g_{ij} c_j 
 +\,  c_j U^g_{ji} \bar c_i -  c_i \bar c_i  U^g_{ji} \big) \; , \nonumber
\end{eqnarray}
so that the difference ({\it i}) - ({\it ii}) yields
\begin{eqnarray}
\sum_i s \big( \bar c_i^a  \, F_i^a \big) &=&  
-  \smallfrac{1}{2\rho} \, \sum_i \sum_{j\sim i}\, \mbox{tr}\,\big(   
 b_i  (  U^g_{ij} -   U^g_{ji}) \big) \label{sMFP}\\
&& \hskip -2.5cm
+\smallfrac{1}{2\rho} \sum_i \sum_{j\sim i}\, \mbox{tr}\,\big(   
\bar c_i  U^g_{ij} c_j   
 -   c_i U^g_{ij} \bar c_j -  [\bar c_i,\, c_i ] \,  \smallfrac{1}{2} (  U^g_{ij} +   U^g_{ji} ) 
 \big) \nonumber \\
&\equiv & \sum_i b_i^a F_i^a \, + \, \sum_{i,\,j} \bar c_i^a \,
     {M_{\mbox{\tiny FP}}}^{ab}_{ij} \,  c_j^b  
 \; , \nonumber
\end{eqnarray}
which defines the lattice Faddeev-Popov operator $M_{\mbox{\tiny FP}}$
of the ghost/anti-ghost symmetric Curci-Ferrari gauges.

Note that the terms in (\ref{stdFP}) can be written in the form
\begin{eqnarray}
\sum_i \bar c_i^a  \, \big( s F_i^a )  &=& \label{stdFP_2} \\
&& \hskip -1.2cm
\smallfrac{1}{4\rho} \, \sum_{i,\,j\sim i}\, \bar c_i^a \, \Big\{ \,
\mbox{tr}\,\big(   [X^a,X^b] (  U^g_{ij} -  U^g_{ji} ) \big) \, (c^b_i
+ c^b_j )  \nonumber\\
&& 
 +\, \mbox{tr}\,\big(   \{X^a,X^b\} (  U^g_{ij} + U^g_{ji} ) \big) \, (c^b_i
- c^b_j )   \, \Big\}  \; . \nonumber
\end{eqnarray}
This yields the standard Faddeev-Popov operator of the 
linear covariant gauges on the lattice. It
differs by the quadratic ghost terms in (\ref{var1}) from the
ghost/anti-ghost symmetric one, $M_{\mbox{\tiny FP}}$  in
(\ref{sMFP}). These terms are of the form
\begin{eqnarray}
- \smallfrac{1}{4\rho} \, \sum_i \sum_{j\sim i}\, \mbox{tr}\,\big(   
 \{ \bar c_i,\, c_i\}\,  (  U^g_{ij} -   U^g_{ji}) \big) &=&\\
  && \hskip -2cm 
\smallfrac{1}{2} \, \sum_i \, \bar c_i^a f^{abc} F_i^c(U^g) \, c_i^b
\;. \nonumber
\end{eqnarray}
In lattice Landau gauge, where $ F_i^a(U^g) = 0$, the forms 
from Eq.~(\ref{stdFP}) and (\ref{sMFP}) both, of course, lead to the same 
Faddeev-Popov operator, as they do in the continuum where the standard
and symmetric versions differ by an analogous term $\propto f^{abc}
\partial_\mu A_\mu^c/2$ which vanishes for $\xi = 0$. For $\xi\not=0$,
on the other hand, the two  Faddeev-Popov operators do
differ and, in particular, only the ghost/anti-ghost 
symmetric framework based on (\ref{sMFP}) leads to a Hermitian one which
can be written in the alternative form, 
\begin{eqnarray}
  \sum_{i,\,j} \bar c_i^a \,
     {M_{\mbox{\tiny FP}}}^{ab}_{ij} \,  c_j^b &=&
     - \smallfrac{1}{4\rho} \sum_{x,\, \mu} \, \Big\{ \label{edBRSFP} \\
&& \hskip -1cm   
 \mbox{tr} \big( \{X^a,X^b\}  (
     U^g_{x,\,\mu} + U^{g\,\dagger}_{x,\,\mu} )\big) \,\times
     \nonumber \\
&& 
  (\bar c_{x+\hat \mu}^a - \bar c_x^a ) (c_{x+\hat \mu}^b - c_x^b )  
 \nonumber \\[2pt]
&& \hskip -1cm   
 + \,  \mbox{tr} \big( [X^a,X^b]  (
     U^g_{x,\,\mu} - U^{g\,\dagger}_{x,\,\mu} )\big) \,\times
     \nonumber \\
&& 
  \big(\bar c_{x}^a  (c_{x+\hat \mu}^b - c_x^b )  -    
     (\bar c_{x+\hat \mu}^a - \bar c_x^a )  c_x^b \big)
   \Big\} \; . \nonumber 
\end{eqnarray}
We have added and subtracted the term proportional
to $\bar c_x^a c_x^b$ in the last line 
here to underpin that in the continuum limit
the $M_{\mbox{\tiny FP}}$ herein reduces to the ghost/anti-ghost
symmetric Faddeev-Popov operator, 
\begin{equation} 
 {M_{\mbox{\tiny FP}}}^{ab}_{ij} \, \stackrel{a\to 0}{\longrightarrow} 
 \,  - a^2\smallfrac{1}{2} \,\big(\partial  D^{ab} + D^{ab}
 \partial \big) \, \delta(x-y)  \, +\, \mathcal{O}(a^4) \; .\nonumber
\end{equation}
To complete the derivation of the gauge-fixing action in the
ghost/anti-ghost symmetric framework, we furthermore need work out the
BRST variation of $s \bar s (\bar c^a c^a) = s (\bar c^a b^a)$ from
(\ref{BRSc})-(\ref{BRSb}). This, however, is done in exactly the same
away as in the continuum, the result is ({\it iii}),
\begin{equation}
  s \big( \bar c^a b^a \big) \, = \, b^a b^a \, +\, \smallfrac{1}{4} 
\, (\bar c \times c )^2 \; . \label{scbb}
\end{equation}
Putting together all terms from ({\it i}) to ({\it iii}) we obtain the
full gauge-fixing action with extended double BRST invariance on the
lattice in the form,
\begin{eqnarray}
S_{\mbox{\tiny GF}} &=& \sum_i \,  \Big\{ - i b^a_i  F_i^a(U^g)
\, - {i}\,\bar c^a_i  {M_{\mbox{\tiny FP}}}_i^a[c]  \label{SGFexpl_app}\\
&& \hskip 2cm +\, 
\smallfrac{\xi}{2} b_i^a b_i^a \, + \, \smallfrac{\xi}{8}\, 
(\bar c_i \times c_i)^2 \, \Big\} \; , \nonumber
\end{eqnarray}
where we introduced the short-hand notation that
\begin{eqnarray}
{M_{\mbox{\tiny FP}}}_i^a[c] &\equiv &  \, -\smallfrac{1}{4\rho}       
 \, \sum_{j\sim i}\,  \Big\{ \,
\mbox{tr}\,\big(   [X^a,X^b] (  U^g_{ij} -  U^g_{ji} ) \big) \, c^b_j
 \nonumber\\ 
&& \hskip .5cm
 +\, \mbox{tr}\,\big(   \{X^a,X^b\} (  U^g_{ij} + U^g_{ji} ) \big) \, (c^b_i
- c^b_j )   \, \Big\}  \; , \nonumber
\end{eqnarray}
which corresponds to the ghost/anti-ghost symmetric
Faddeev-Popov operator in (\ref{edBRSFP}), in particular, we have
\begin{equation}
 \sum_{i} \bar c_i^a \,
     {M_{\mbox{\tiny FP}}}^{a}_{i}[c]  \, = \, 
 \sum_{i,\,j} \bar c_i^a \,
     {M_{\mbox{\tiny FP}}}^{ab}_{ij} \,  c_j^b \, ,
\end{equation}
with ${M_{\mbox{\tiny FP}}}^{ab}_{ij}$ in a simplified alternative
form given by
\begin{eqnarray}
{M_{\mbox{\tiny FP}}}_{ij}^{ab} &= &  \, -\smallfrac{1}{2\rho}       
 \, \sum_{k\sim i}\,  \Big\{ \,
\mbox{Re tr}\,\big( \{X^a,X^b\} \, U^g_{ik} \big) \, \delta_{ij}
\label{sBRSTFP} \\ 
&& \hskip 2.5cm
 -\,2\, \mbox{Re tr}\,\big( X^b X^a \,  U^g_{ik} \big) \, \delta_{kj} 
 \, \Big\}  \; . \nonumber
\end{eqnarray}
The Faddeev-Popov operator of lattice Landau gauge was first derived
in \cite{Zwanziger1994}. It might be worth recalling that the derivation
presented there, based on the differentials of the gauge-fixing potential 
$V_U[g]$ along one-parameter subgroups of the $SU(N)$ gauge group, by
construction leads directly to the Hermitian $M_{\mbox{\tiny  FP}}$ in
(\ref{sBRSTFP}), and not to that of standard Faddeev-Popov theory on the
lattice following from (\ref{stdFP_2}). They are equivalent in Landau gauge,
of course. Their subtle difference needs to be kept in mind, however,
when attempting to implement smeared covariant gauges for $\xi\not=0$
on the lattice as was done, {\it e.g.}, in
Refs.~\cite{Giusti1996,Giusti1999}. It reflects the different symmetry
properties of standard Faddeev-Popov theory and the ghost/anti-ghost symmetric 
framework for $\xi\not=0$.

\section{Expectation values in the `would-be-Gauss-Bonnet' limit} 
\label{expl_calcs}

Here we describe the explicit calculations to derive the results for
the expectation values in Eq.~(\ref{aux_3}), as summarized in
Table~\ref{res_tab}, which are needed at the $2^\mathrm{nd}$ order in
the expansion of the Curci-Ferrari mass parameter $\widehat m^2(x)$
around $x=0$, the 'would-be-Gauss-Bonnet' limit.  

For the first term we use $\langle b_i^a b_j^b \rangle_{\widehat
  m^2\!,\,0} =
\delta^{ab} \delta_{ij}$ to obtain,
\begin{equation} 
 \big\langle\,  \big(i b,F\big)^2  \big\rangle_{\widehat m^2\!,\,0}
 =\, - \ \big\langle\,  \big(F,F\big)\,  \big\rangle_{\widehat
 m^2\!,\,0} \, .
\end{equation}
Remember the explicit form (\ref{std-gfc}) of the gauge condition,
\begin{equation} 
 F^a_i(U^g) \, = \,  -  \sum_{j\sim i}
 \mbox{tr}\,\big( X^a(  g_i^\dagger U_{ij} g_j  -   g_j^\dagger U_{ji}
 g_i ) \big) \; . \label{aux_gc}
\end{equation}
All terms in the sum of the squares of this condition are quadratic in
$g$ and in $g^\dagger$. For $N\ge 3$ the only non-vanishing results of
the group integration arise in terms where all the $g_i$'s 
match up pairwise with $g_i^\dagger $'s at identical sites (the
special case of $N=2$ will be discussed separately below). 
We can then use for the fundamental $g$'s at that site $i$,
\begin{equation}
  V_N^{-1} \int dg_i  \, (g_i)_{kl} \, (g^\dagger_i)_{mn} \, = \,
  \smallfrac{1}{N}   \, \delta_{kn} \delta_{lm} \; . \label{invint_1}
\end{equation} 
In one particular term $F_i^a F_i^a$, without summation over $i$, only the
two mixed terms then survive and we have,
\begin{eqnarray}
V_N^{-1}\! \int dg_i  \, F_i^a F_i^a  &=& -2 V_N^{-1}\! \int
  dg_i  \, \times  \hspace{1.2cm}  \mbox{(no sum $i$)} \nonumber \\
 && \hskip -1.2cm  
\Big(\sum_{j\sim i} \mbox{tr} \,\big( X^a  g_i^\dagger
  U_{ij} g_j \big) \Big)
    \Big(\sum_{k\sim i} \mbox{tr} \,\big( X^a  g_k^\dagger
  U_{ki} g_i \big) \Big) \nonumber\\
&& \hskip -.8cm =  -\smallfrac{2}{N} \sum_{k,j\sim i} \mbox{tr}
  \,\big( X^aX^a  g_k^\dagger 
  U_{ki}  U_{ij} g_j \big) \nonumber\\
  &&  \hskip -.8cm = 
 \left(1 -\smallfrac{1}{N^2} \right)  \sum_{k,j\sim i} \mbox{tr}
  \,\big( g_k^\dagger   U_{ki}  U_{ij} g_j \big)\; . \label{aux_4}
\end{eqnarray}
The integrations over the gauge group elements at all sites $k\sim i$,
attached to site $i$, by the same argument yield a non-vanishing
result only if $j=k$ in the double sum over all neighbors of site $i$
in (\ref{aux_4}). For those terms the group integration yields 
$\mbox{tr}\, U_{ji} U_{ij} = N$. Therefore, in $D$ dimensions, 
\begin{equation}
 \big\langle\,  \big(F,F\big)\,  \big\rangle_{\widehat
 m^2\!,\,0} = 
 \mathcal{N} \! \int \prod_{i} dg_i \, \big(F,F\big)  =
  \, (\# \mbox{\small sites}) \times 4 D \, C_2^f \, , 
\end{equation}
for $SU(N)$ with $N\ge 3$, where $C_2^f = \smallfrac{1}{2} \big( N -
1/N \big)$ is the value of the quadratic Casimir operator in the
fundamental representation. 

For $N=2$ we obtain an additional contribution to $ \langle\,
(F,F)\,  \rangle_{\widehat m^2\!,\,0}$ from   
the squares of the two terms in the gauge condition (\ref{aux_gc}). 
This is because for $g \in SU(2)$,
\begin{equation} 
 V_2^{-1} \int dg_i \, (g_i)_{kl} \, (g_i)_{mn} \, = \,
 \smallfrac{1}{2} \, \epsilon_{km} \epsilon_{ln} \; . \label{invint_2}
\end{equation}
Again, however, only the squares of the same links contribute in the
double sum over the neighbors of site $i$. For these, {\it e.g.},
\begin{equation}
  \left(\mbox{tr}\, X^a g_j^\dagger U_{ji} g_i \right)^2\,
  \stackrel{V_2^{-2} \int dg_i dg_j}{\longrightarrow}\;  \smallfrac{3}{4}
  \, \mbox{det} \, U_{ji} = C_2^f \; . 
\end{equation}
There are two such terms  in $F_i^a F_i^a$ for each of the $2D$ links
attached to site  $i$. 
Therefore, the total additional contribution from those terms in $SU(2)$ equals
that from the mixed terms calculated for general $N$ above, we have,  
\begin{equation}
 \big\langle\,  \big(F,F\big)\,  \big\rangle_{\widehat
 m^2\!,\,0} = \, 
\, (\# \mbox{\small sites}) \times 8 D \, C_2^f 
\; , \;\; \mbox{for $SU(2)$\,.}
\end{equation}
Next, to compute the expectation value $\propto \langle\, V_U^2\,
  \rangle_{\widehat  m^2\!,\,0} $ of 
\begin{equation} 
V_U^2[g] \, = \, \Big( \sum_{i,\, j\sim i} \mbox{tr}\,
  g_i^\dagger U_{ij} g_j  \Big)\, \Big( \sum_{k, \, l\sim k} \mbox{tr}\,
  g_k^\dagger U_{kl} g_l  \Big) \; , \label{aux_V2}
\end{equation}
we consider one term for fixed (neighbors) $i$ and $j$ in the double
sum. Integrating this term over $g_j$ via (\ref{invint_1}),
\begin{eqnarray} 
V_N^{-1}\!\int dg_j \; \mbox{tr}\,
  g_i^\dagger U_{ij} g_j  \, \Big( \sum_{k, \, l\sim k} \mbox{tr}\,
  g_k^\dagger U_{kl} g_l  \Big) &=&\hskip 2cm \label{aux_5}\\
&& \hskip -2cm 
\smallfrac{1}{N} \sum_{l\sim j} \, 
 \mbox{tr}\,
  g_i^\dagger U_{ij} U_{jl} g_l \; , \nonumber
\end{eqnarray}
for $N\ge 3$, where the only contribution arises when $k=j$. Then, $l$ must
be one of the neighbors of $j$. Successive integration over $g_i$
singles out that neighbor of $j$ with $l=i$,
\[
V_N^{-1}\!\int dg_i \,
\smallfrac{1}{N} \sum_{l\sim j} \, 
 \mbox{tr}\,
  g_i^\dagger U_{ij} U_{jl} g_l \, = \, \smallfrac{1}{N} \,
  \mbox{tr}\, U_{ij} U_{ji} \, = \, 1 \; . \]
We obtain one such contribution for every one of the $2D$ neighbors 
$j\sim i$ at site $i$, thus summing $\sum_{i,j\sim i} $ yields
\begin{equation} 
\langle\, V_U^2[g]\, \rangle_{\widehat
  m^2\!,\,0} \, = \, (\# \mbox{\small sites}) \times 2D \;, \;\; N\ge 3 \; .
\label{V2_Nge3}\end{equation}
Note that the number of sites in all these expectation values cancels
with that in (\ref{aux_2}) when computing the ratio of Eq.~(\ref{mhat-der}). 

Again, for $N=2$ in $SU(2)$ we obtain an additional contribution from
(\ref{invint_2}). Starting again from the contribution to (\ref{aux_V2})
for fixed neighbors $i$ and $j$  as in (\ref{aux_5}) we now obtain
for $N=2$,  
\begin{eqnarray} 
V_2^{-1}\!\int dg_j \; \mbox{tr}\,
  g_i^\dagger U_{ij} g_j  \, \Big( \sum_{k, \, l\sim k} \mbox{tr}\,
  g_k^\dagger U_{kl} g_l  \Big) &=&  \hskip 2cm \label{aux_6}\\
&& \hskip -6.5cm 
\smallfrac{1}{2} \sum_{l\sim j} \, 
 \mbox{tr}\,
  g_i^\dagger U_{ij} U_{jl} g_l \, + 
 \smallfrac{1}{2} \sum_{k\sim j} 
 \epsilon_{su} \, \epsilon_{rt} \,
  (g_i^\dagger U_{ij})_{sr} (g_k^\dagger U_{kj})_{ut} 
\; . \nonumber
\end{eqnarray}
The second group integration over $g_i$ then produces, in
addition to the above, an according contribution from the second term,
which is given by
\begin{eqnarray} 
V_2^{-1}\!\int dg_i \;  \smallfrac{1}{2} \sum_{k\sim j} 
 \epsilon_{su} \, \epsilon_{rt} \,
  (g_i^\dagger U_{ij})_{sr} (g_k^\dagger U_{kj})_{ut} 
   &=&  \hskip 1cm \label{aux_7}\\
&& \hskip -6.5cm 
\smallfrac{1}{4} \,  \epsilon_{su} \, \epsilon_{rt} \, 
 \epsilon_{su} \, \epsilon_{vw} \, 
  (U_{ij})_{vr} (U_{ij})_{wt} \, =\,  \smallfrac{1}{4} \, 4 \,
 \mbox{det} \, U_{ij} \, = \, 1
\; . \nonumber
\end{eqnarray}
This equals the first term obtained for all $N$; and 
together the two again give for $SU(2)$ twice the 
result for $N\ge 3$ in (\ref{V2_Nge3}) above, {\it i.e.},
\begin{equation} 
\langle\, V_U^2[g]\, \rangle_{\widehat
  m^2\!,\,0} \, = \, (\# \mbox{\small sites}) \times 4D \;, \;\; N=2 \; .
\end{equation}

The hardest task here is to compute the last remaining term in
(\ref{aux_3}), $  \big\langle\,  \big( i\bar c ,
  M_{\mbox{\tiny FP}}c\big)^2  \big\rangle_{\widehat m^2\!,\,0} $. For
  a start, we first note that
\[
\langle i\bar c_i^a c_j^b \,  i\bar c_k^c c_l^d  \big\rangle_{\widehat
  m^2\!,\,0} = \left( \delta^{ab} \delta_{ij} \, \delta^{cd}
  \delta_{kl} - \, \delta^{ad} \delta_{il} \, \delta^{bc} \delta_{jk}
  \right) \, A^{ac}_{ik} \; , 
\]
where 
\begin{equation}
 A^{ab}_{ij} \,=\,  \langle (i\bar c\, c)_i^a \,  (i\bar c\, c)_j^b
  \big\rangle_{\widehat 
  m^2\!,\,0}  \; , \label{aux_18}
\end{equation}
using the notation  $(i\bar c\, c)_i^a \equiv i\bar c_i^a  c_i^a $
without implicit summations over $a$ and $i$, here. The expectation
value in (\ref{aux_18}) is of course independent of $\{U\}$. It 
depends on the site indices only in that we
need to distinguish whether $i=j$ or not, {\it i.e.}, 
\begin{equation}
 A^{ab}_{ij} =  \left\{ 
\begin{array}{ll}
P_N^{ab}(\widehat m_0^2)\,, \; 
P_N^{ab}=\left\{ \begin{array}{cl}
 P_N^{ba}\, , & \! a\not=b\\
  0 \, , & \! a=b   \end{array}\right.
\, ; &  i=j \; .\\[12pt]
Q_N(\widehat m_0^2)\, , \;\; \mbox{independent of } \{a,\, b\} \, ; &
 i\not=j \, .
\end{array}
  \right.
  \label{aux_19}
\end{equation}
where both $P_N^{ab}$ and $Q_N$ are site-independent.
We have
\begin{equation}
Q_N(\widehat m_0^2)\,= \frac{1}{(N^2-1)^2}\,
  \left(\frac{I_N'(\widehat m_0^2)}{I_N(\widehat 
  m_0^2)}\right)^2 \; , \label{aux_20}
\end{equation}
from (\ref{aux_1}), while $P_N^{ab}$ has a more complicated structure, in
general. Only its sum simplifies,
\begin{equation}
\sum_{a,\, b} P_N^{ab}(\widehat m_0^2)\, = \, 2 \sum_{a,\, b>a}
P_N^{ab}(\widehat m_0^2)\, = \,  \frac{I_N''(\widehat m_0^2)}{I_N(\widehat
  m_0^2)}  \; . \label{aux_21}
\end{equation}
which follows immediately from its definition via (\ref{aux_18}) and
(\ref{aux_19}) and with Eqs.~(\ref{qGaussBonnet})--(\ref{cbc-siteint}). 
With these results,
\begin{eqnarray}
 \big\langle\,  \big( i\bar c ,
  M_{\mbox{\tiny FP}}c\big)^2  \big\rangle_{\widehat m^2\!,\,0} &=&
\label{aux_22}\\
&& \hskip -2cm
\big\langle \!\sum_{i,j,a,b} \big( M^{aa}_{ii} M^{bb}_{jj} - 
  M^{ab}_{ij} M^{ba}_{ji}  \big) \, A^{ab}_{ij} \, \big\rangle_{\widehat
  m^2\!,\,0} \nonumber \\
&& \hskip -3.5cm
= \,2Q_N(\widehat m_0^2 ) \!\!\!
  \sum_{i,j>i,a,b} \big\langle  \big( M^{aa}_{ii} M^{bb}_{jj} - 
  M^{ab}_{ij} M^{ba}_{ji}  \big) \, \big\rangle_{\widehat
  m^2\!,\,0}  \nonumber \\
&& \hskip -3cm
+ \, 2\!\!\!\sum_{i,a,b>a} P_N^{ab}(\widehat m_0^2) \, 
     \big\langle  \big( M^{aa}_{ii} M^{bb}_{ii} - 
  M^{ab}_{ii} M^{ba}_{ii}  \big) \, \big\rangle_{\widehat
  m^2\!,\,0} \; .   \nonumber 
\end{eqnarray}
The first part with the contributions from different sites $j\not=i$
would be a disaster for the intended $\widehat m_0^2 \to 0$ limit: 
Because $I'_N(\widehat m_0^2)/I_N(\widehat m_0^2) $ always is of the order
$1/\widehat m_0^2$, we have that $Q_N(\widehat m_0^2) $ is of the order 
$1/\widehat m_0^4$. Recalling that we need to divide all terms computed
here by the expectation value in (\ref{aux_2}) which is proportional to
$I_N'(\widehat m_0^2)/I_N(\widehat m_0^2) $, this then implies that
the second derivative w.r.t.~$x$ of $\widehat m^2 (x)$ would contain a
contribution proportional $1/\widehat m_0^2$ at $x=0$ and therefore
become infinite in the limit $\widehat m_0^2 \to 0$. Luckily, this
contribution turns out to be zero for all $N$ because of a
cancellation between the two terms in the expectation value of this
part in (\ref{aux_22}). To see this, first consider, 
\begin{eqnarray}
\sum_{a,b} M^{aa}_{ii} M^{bb}_{jj}  &=& \label{aux_23} \\
&& \hskip -1cm 
4{C_2^f}^2 
\sum_{k\sim  i}\sum_{l\sim j}    \, 
\mbox{Re tr}\,\big(g_i^\dagger U_{ik} g_k \big) \,
  \mbox{Re tr}\,\big(g_j^\dagger U_{jl} g_l\big)
\nonumber \\
&& \hskip -1.4cm  = \, {C_2^f}^2 
\sum_{k\sim  i}\sum_{l\sim j}    \, 
\Big( \mbox{tr}\, U_{ik}^g  \, \mbox{tr}\, U_{jl}^g\, + \, 
\mbox{tr}\, U_{ki}^g  \, \mbox{tr}\, U_{lj}^g\, + \nonumber \\
&& \hskip 1.4cm
\mbox{tr}\, U_{ik}^g  \, \mbox{tr}\, U_{lj}^g\, + \, 
\mbox{tr}\, U_{ki}^g  \, \mbox{tr}\, U_{jl}^g
\Big)  \;. \nonumber 
\end{eqnarray}
Because $i\not=j$, only the first two terms in the brackets contribute
when integrating $g_i$ via (\ref{invint_1}). For those, the different group
integrations over $g_i$ and $g_j$, by the method now familiar, then
yield,
\begin{equation} 
V_N^{-2} \int dg_i \, dg_j \,  \sum_{a,b} M^{aa}_{ii} M^{bb}_{jj}  \,
=\,   2{C_2^f}^2 \sum_{k\sim  i}\sum_{l\sim j}    \, \delta_{li} \, 
\delta_{kj} \; . 
\end{equation}
After summation over $j$ we can replace $j$ by $k$ here. The fact that
in this sum we need to restrict $j\not= i$ does not matter because 
the non-zero contributions arise for $j=k$, where $k $ is a neighbor
of $i$ and thus necessarily different from $i$. Subsequent summation
over the neighbors $l$ of $k$ now, then due to the second Kronecker 
symbol picks up neighbor $i$ of site $k$. We have
\begin{eqnarray} 
\sum_{i,j\not=i} V_N^{-2} \int dg_i \, dg_j \,  \sum_{a,b} M^{aa}_{ii}
M^{bb}_{jj}  &=&   2{C_2^f}^2 \sum_{i, k\sim  i} \sum_{l\sim k} \,
\delta_{li}  \nonumber \\
&& \hskip -3cm =\,  (\# \mbox{\small sites}) \times D \,
\left(N-\smallfrac{1}{N} 
\right)^2 \; . \label{aux_25} 
\end{eqnarray}
As before there is an additional contribution from (\ref{invint_2})
for $SU(2)$ here also. As $i$ and $j$ are different, after
integrating $g_i$ this time, this contribution is nonzero only for
precisely the other two terms in the brackets above. Using
(\ref{invint_2}) it is straightforward to verify that the result for
those two terms again matches the contribution just calculated from
(\ref{invint_1}). We can therefore summarize that for all $N$ including
$N=2$,   
\begin{equation} 
\sum_{i,j\not=i,a,b}  \big\langle M^{aa}_{ii}
M^{bb}_{jj}  \big\rangle_{\widehat
  m^2\!,\,0}  =\,  (\# \mbox{\small sites}) \times 2D \,
\, 2 {C_2^f}^2 
\big( 1  +  \delta_{N,2} \big) \; . \label{aux_26} 
\end{equation}
For the second term in the first part of (\ref{aux_22}) with $i\not=
j$, which allowed us to drop the diagonal terms in both Faddeev-Popov
operators, we need that in the fundamental representation, 
\[
(X^bX^a)_{ij} (X^aX^b)_{kl} \, =\, \smallfrac{1}{4} \left(
  N-\smallfrac{2}{N}\right) \delta_{il} \delta_{jk} +
  \smallfrac{1}{4N^2} \, \delta_{ij}\delta_{kl} \; ,
\]
and
\[
(X^aX^b)_{ij} (X^aX^b)_{kl} \, =\, - \smallfrac{1}{2N} \,
  \delta_{il} \delta_{jk} +
  \smallfrac{1}{4}  \left(
  1+\smallfrac{1}{N^2}\right) \delta_{ij}\delta_{kl} \; ,
\]
so that we can write (for $i\not=j$),
\begin{widetext}
\begin{eqnarray}
   \sum_{a,b} M^{ab}_{ij} M^{ba}_{ji}  &=& 
4 \sum_{k\sim  i}\sum_{l\sim j}    \, 
\mbox{Re tr}\,\big(X^bX^a g_i^\dagger U_{ik} g_k \big) \,
\mbox{Re tr}\,\big(X^a X^b g_j^\dagger U_{jl}
g_l\big) \, \delta_{kj} \delta_{li} \label{aux_27} \\
&=& 
 \sum_{k\sim  i}\sum_{l\sim j}    \, 
\Big( \mbox{tr}\,X^bX^a U_{ij}^g  \,+ \, \mbox{tr}\,X^aX^b U_{ji}^g \Big) 
\Big( \mbox{tr}\,X^aX^b U_{ji}^g  \, + \, 
\mbox{tr}\,X^bX^a U_{ij}^g \Big) \,
  \delta_{kj} \delta_{li} \; = \nonumber\\
&& \hskip -2.5cm  
\sum_{k\sim  i}\sum_{l\sim j}    \, 
\Big\{ \smallfrac{1}{2} \left(N^2-2\right) 
\,+ \, \smallfrac{1}{2N^2} \,  \mbox{tr}\, U_{ij}^g \,  
                                     \mbox{tr}\,  U_{ji}^g  \, 
-  \smallfrac{1}{2N} \,\Big(  \mbox{tr}\, U_{ij}^g U_{ij}^g +
\mbox{tr}\, U_{ji}^g  U_{ji}^g \Big) \, +\,
  \smallfrac{1}{4}  \left(
  1+\smallfrac{1}{N^2}\right)   \Big( \big(\mbox{tr}\, U_{ij}^g\big)^2
+ \big(\mbox{tr}\, U_{ji}^g\big)^2 \Big)\Big\}\,
  \delta_{kj} \delta_{li} 
  \;. \nonumber 
\end{eqnarray}
\end{widetext}
The group integrations via (\ref{invint_1}) and, in addition,  
for the special case of $N=2$ via (\ref{invint_2}) proceed as before.
The explicit results of the corresponding calculations above can 
all be reduced to essentially using two relations summarized as follows,
\begin{eqnarray}
&& \hskip -.8cm 
\big\langle \mbox{tr}\,\big(g_i^\dagger U_{ij} g_j \big)
 \mbox{tr}\,\big( g_k^\dagger U_{kl} g_l 
 \big) \big\rangle_{\widehat
  m^2\!,\,0} =\, \delta_{il} \delta_{ki} \, + \, \delta_{N,2}
 \, \delta_{ik} \delta_{jl} \; ,\nonumber \\  
&& \hskip -.8cm 
\big\langle \mbox{tr}\,\big(g_i^\dagger U_{ij} g_j g_k^\dagger U_{kl} g_l 
 \big) \big\rangle_{\widehat
  m^2\!,\,0} =\, N\, \delta_{il} \delta_{ki} \, - \, \delta_{N,2}
 \, \delta_{ik} \delta_{jl} \; .\nonumber \\
\label{aux_28}
\end{eqnarray}
These are the basic terms that arise at the quadratic order in $x$,
and hence in the gauge-transformed links $U^g$, of
our mass expansion (recall that the terms linear in $x$ vanish in this
expansion because the linear order terms in $U^g$ do upon the group
integrations).  Using the relations (\ref{aux_28}) in (\ref{aux_27}),
we obtain, 
\begin{widetext}
\begin{eqnarray}
\sum_{a,b} 
\big\langle M^{ab}_{ij} M^{ba}_{ji} \big\rangle_{\widehat
  m^2\!,\,0} \, = \, \left\{
 \smallfrac{1}{2} \left(N^2-2\right) 
\,+ \, \smallfrac{1}{2N^2} \, + \, \delta_{N,2} \,\left(
  \smallfrac{1}{N} \, +\,  \smallfrac{1}{2}  \left(
  1+\smallfrac{1}{N^2}\right) \right) \right\} \, \sum_{k\sim
    i}\sum_{l\sim j}  \delta_{kj} \delta_{li} \; ,
 \label{aux_29}
\end{eqnarray}
and hence, together with (\ref{aux_26}),
\vspace{-.4cm}
\begin{equation}
\sum_{i,j\not= i,a,b} \big\langle  \big( M^{aa}_{ii} M^{bb}_{jj} - 
  M^{ab}_{ij} M^{ba}_{ji}  \big) \, \big\rangle_{\widehat
  m^2\!,\,0} \, =  \, 
 (\# \mbox{\small sites}) \, 2D \, \times   
\left\{ \begin{array}{ll}
 \smallfrac{1}{2} \left(N-\smallfrac{1}{N}\right)^2 
 - \smallfrac{1}{2} \left(N^2-2\right) 
-  \smallfrac{1}{2N^2} \, = \, 0 \, , & N\ge 3\, ,\\[8pt]
\left(N-\smallfrac{1}{N}\right)^2 
 - \smallfrac{1}{2} \left(N^2-2\right) 
-  \smallfrac{1}{2N^2} & \\[4pt] 
 \hskip 1.8cm - \left(\smallfrac{1}{N} +
  \smallfrac{1}{2}\left(1+\smallfrac{1}{N^2}\right)\right) 
\, = \, 0\, , & N=2 \, . 
	\end{array} \right. \label{aux_30} 
\end{equation}
\end{widetext}
The two terms from (\ref{aux_26}) and (\ref{aux_29}) therefore cancel
and the first part in (\ref{aux_22}) thus vanishes in either case,
whether $N=2$ or $N\ge 3$.  At the same time this cancels the
otherwise quite disastrous singularity of the mass expansion in the
$\widehat m_0^2 \to 0$ limit, as promised. 

In the second term in (\ref{aux_22}), the one with $i=j$, we need
products of diagonal entries of the Faddeev-Popov operator of the form
({\it c.f.}, Eq.~(\ref{sBRSTFP}); no implicit sum over $i$ here either),
\begin{eqnarray}
M_{ii}^{ab} 
&=&  - \sum_{k\sim i}   \smallfrac{1}{2} \left( 
\mbox{tr}\,\{X^a, X^b\}\, U_{ik}^g \, +\,   \mbox{tr}\,\{X^a, X^b\}\,
U_{ki}^g \right) \nonumber  \\
&=&   \sum_{k\sim i} \left\{  \smallfrac{1}{2N^2}\, \delta^{ab} \big( 
\mbox{tr}\, U_{ik}^g \, +\,   \mbox{tr}\,U_{ki}^g \big)
\right.   \label{aux_31} \\ 
&&\hskip 1cm 
\left. +\, \smallfrac{1}{2}\, d^{abc} \big( 
\mbox{tr}\,iX^c U_{ik}^g \, +\,   \mbox{tr}\, iX^cU_{ki}^g \big)
\right\}
\; .     \nonumber
\end{eqnarray}
Where we have used the identity
\begin{equation}
\{X^a, X^b\} \, =\, -\smallfrac{1}{N}\, \delta^{ab}- i d^{abc} X^c\; . 
\end{equation}
For $SU(2)$ we set  $ d^{abc} = 0$ which then leaves us with only the first
term in (\ref{aux_31}).
Because, {\it e.g.},
\begin{eqnarray} 
 \big\langle \mbox{tr}\,\big( g_i^\dagger U_{ik} g_k \big)
 \, \mbox{tr}\,\big(iX^a g_l^\dagger U_{li} g_i \big)  
\big\rangle_{\widehat
  m^2\!,\,0} &=& \label{aux_33} \\
\smallfrac{1}{N} \,  \big\langle \mbox{tr}\,\big(iX^a g_l^\dagger
 U_{li} U_{ik} g_k \big)   \big\rangle_{\widehat
  m^2\!,\,0} &=& 0 \; ,\nonumber
\end{eqnarray}
due to the tracelessness of the generators $X^a$, there will be no
terms linear in the $X$'s in the expectation values of squares of the
operators in (\ref{aux_31}) for $N\ge 3$ either. The terms quadratic
in the $X$'s simplify from relations as follows,
\begin{eqnarray} 
 \big\langle \mbox{tr}\,\big(iX^a g_i^\dagger U_{ik} g_k \big)
 \, \mbox{tr}\,\big(iX^b g_l^\dagger U_{li} g_i \big)  
\big\rangle_{\widehat
  m^2\!,\,0} &=&   \label{aux_34} \\
 \smallfrac{1}{N} \,
  \big\langle \mbox{tr}\,\big(iX^b g_l^\dagger U_{li} U_{ik} g_k iX^a\big)  
\big\rangle_{\widehat
  m^2\!,\,0} &=& \smallfrac{1}{2N} \, \delta^{ab} \, \delta_{kl} 
\; .\nonumber
\end{eqnarray}
We do not have to worry about corresponding terms from
(\ref{invint_2}) for $N=2$ here because these contain the $d$-symbols
  which are zero in $SU(2)$. With (\ref{aux_28}), (\ref{aux_33}) and
  (\ref{aux_34}) we therefore obtain,
\begin{align}
\big\langle  M^{ab}_{ii} M^{cd}_{ii} \, \big\rangle_{\widehat
  m^2\!,\,0} &= \\
& \hskip -1cm \sum_{k\sim i} \left\{ \smallfrac{1}{2N^2}
\, \delta^{ab} \delta^{cd} \, \big(1 + \delta_{N,2} \big) 
\, + \smallfrac{1}{4N} \, d^{abe} d^{cde} 
\right\} \nonumber\\
& \hskip -.2cm 
\stackrel{N>2}{=} \, \smallfrac{2D}{4N} \, \left(\smallfrac{2}{N}
 \, \delta^{ab} \delta^{cd} \, + \, d^{abe} d^{cde} 
\right)\; . \nonumber
\end{align}
We therefore have, implicitly summing over color indices again
 but not over sites $i$ (yet), 
\begin{align}
\big\langle (i\bar c^a M^{ab} c^b)_i \,  (i\bar c^c M^{cd} c^d)_i 
  \big\rangle_{\widehat  m^2\!,\,0} &= \, \label{aux_36}\\ 
\big\langle (i\bar c^a c^b)_i \,  (i\bar c^c c^d)_i 
  \big\rangle_{\widehat  m^2\!,\,0}
  & \big\langle  M^{ab}_{ii} M^{cd}_{ii} \, \big\rangle_{\widehat
  m^2\!,\,0} \nonumber \, = \, \nonumber \\
& \hskip -4.5cm 
-\smallfrac{2D}{4N} \left\{ \smallfrac{2}{N}  \big\langle (\bar c^ac^a)_i^2 
  \big\rangle_{\widehat  m^2\!,\,0}  \, + \,  d^{abe} d^{cde}
\big\langle (\bar c^a c^b \bar c^c c^d)_i   \big\rangle_{\widehat
  m^2\!,\,0}  \right\}   \nonumber \\
 & \hskip -4.5cm  
=\, -\smallfrac{2D}{4N}
\big\langle  f^{abe} (\bar c^a c^b)_i  \,  f^{cde}
(\bar c^c c^d)_i \big\rangle_{\widehat m^2\!,\,0} 
  \, = \nonumber \\
 & 
- \smallfrac{2D}{4N}  \big\langle (\bar c \times c)_i^2
\big\rangle_{\widehat  m^2\!,\,0} \; , \nonumber 
\end{align}
 for $N\ge 3$, and twice that for $N=2$ again, where the $d$'s are
 zero and where the $x=0$ expectation value of the remaining first
 term above  agrees with that of the quartic ghost interaction,
\begin{equation}
 - \big\langle 
(\bar c \times c)_i^2 \big\rangle_{\widehat  m^2\!,\,0}  \, = 
 \big\langle (i\bar c^ac^a)_i^2 
  \big\rangle_{\widehat  m^2\!,\,0}  \, = \, 
  \frac{I_2''(\widehat m_0^2)}{I_2(\widehat m_0^2)} \; . \label{aux_37}
\end{equation}
Finally, summing in (\ref{aux_36}) over the sites $i$ of the lattice,
and from (\ref{aux_22}) with (\ref{aux_29}), we obtain 
\begin{eqnarray}
 \big\langle\,  \big( i\bar c ,
  M_{\mbox{\tiny FP}}c\big)^2  \big\rangle_{\widehat m^2\!,\,0} &=&
 2D \, (\#
  \mbox{\small sites}) \times 
\label{aux_38}\\
 && \hskip -2.2cm 
\left\{ \begin{array}{ll}
-\smallfrac{1}{4N}\,  \langle (\bar c\times c)^2 \rangle_{\widehat
  m^2\!,\,0} \equiv
\smallfrac{2}{N} \, \smallfrac{J_N(\widehat m_0^2)}{I_N(\widehat
  m_0^2)} \, , & N\ge 3\, , \\[10pt]
- \smallfrac{1}{4}\,\langle (\bar c \cdot c)^2 \rangle_{\widehat m^2\!,\,0} = 
   \smallfrac{1}{4}\, \smallfrac{I_2''(\widehat m_0^2)}{I_2(\widehat m_0^2)} \,
 , & N=2 \, . 
	\end{array} \right. \nonumber
\end{eqnarray}
The only piece left to compute is the expectation value (at $x=0$) of
the quartic ghost interaction,
\begin{equation}
  \label{aux_39}
  -\smallfrac{1}{8} \big\langle (\bar c \times c)^2
  \big\rangle_{\widehat m^2\!,\,0} \, \equiv \, \frac{J_N(\widehat
    m_0^2)}{I_N(\widehat   m_0^2)} \; . 
\end{equation}
The integral expression for $J_N$, analogous to that for $I_N$, {\it
  c.f.}, Eq.~(\ref{cbc-siteint}), is given by
\begin{eqnarray} 
 J_N(\widehat m^2) & = & \int \prod_{a=1}^{N^2\!-1} d(i\bar c^a) d c^a
\, \left(  -\smallfrac{1}{8} (\bar c \times c)^2 \right)
 \, \label{int-siteint}\\
&& \hskip 1.4cm
 \exp\left\{ i \widehat m^2\,  \bar c\! \cdot \! c \, - \,
 \smallfrac{1}{8}\,   (\bar c \times c)^2 \right\} \; . \nonumber
\end{eqnarray}
For $SU(2)$ and  $SU(3)$, respectively, the resulting 
$J_2(\widehat m^2)$ and $J_3(\widehat m^2)$ are given  explicitly in
Eqs.~(\ref{J_2}) and  (\ref{J_3}). 
This completes the computations of the terms in (\ref{aux_3}). The
results are summarized in Table~\ref{res_tab}.

\end{document}